# A framework for evaluating biosignature potential against the abiotic baseline on ocean worlds

Peter M. Higgins[1,2], Weibin Chen[2], Oliver Warr[3], Lucas M. Fifer[4], Wanying Kang[5], Charles S. Cockell[6], and Barbara Sherwood Lollar[2,7]

[1]*Department of Earth and Planetary Sciences, Harvard University, Cambridge, MA, USA*
[2]*Department of Earth Sciences, University of Toronto, Toronto, ON, Canada*
[3]*Department of Earth and Environmental Sciences, University of Ottawa, Ottawa, Ontario, Canada*
[4]*Department of Earth and Space Sciences, University of Washington, Seattle, WA, USA*
[5]*Earth, Atmospheric and Planetary Science Department, Massachusetts Institute of Technology, Cambridge, MA, USA*
[6]*UK Centre for Astrobiology, University of Edinburgh, Edinburgh, UK*
[7]*Institut de Physique du Globe de Paris (IPGP), Université Paris Cité, Paris, France*

*Corresponding author: Peter M. Higgins (phiggins@fas.harvard.edu)*

## Abstract

Ocean worlds are considered as targets for life detection missions because they meet several key requirements for habitability. However, identifying potential life on other worlds requires observing clear and unambiguous biosignature signals above the existing abiotic baseline. Consequently, this necessitates evaluating uncertainty and variability in the abiotic baseline, including processes that can overlap, attenuate, or obfuscate biosignatures before they are observed. This article develops a quantitative framework for holistically evaluating abiotic baselines on ocean worlds to guide life detection strategies. Using Enceladus as an example, we assess the potential of using: i) $CH_4$ isotopes and their relationship with $CO_2$, and ii) amino acid chirality as biosignatures, demonstrating that uncertainties in abiotic processes currently prevent hypothetical future $\delta^{13}C_{CO2}$ and $\delta^{13}C_{CH4}$ measurements from definitively inferring a biosphere on Enceladus. Additionally, our results quantitatively show that neglecting the abiotic baseline risks false negative life detection claims for both isotopic and chiral biosignatures. Interpreting these and other alternative biosignatures on Enceladus, Europa, Titan, and similar planetary bodies therefore requires complimentary geophysical observations such as constraining internal temperatures to within ~10-100°C, and improving characterisation of the target's rheology, lithology, initial abiotic organic inventory and ocean transport timescales.

Icy ocean worlds are common in the solar system and some of them may be habitable[1–6]. Observations from the Cassini-Huygens mission and subsequent modelling suggest that Enceladus may contain habitable areas near its seafloor[3–7], and in the next decade JUICE, Europa Clipper and Dragonfly will quantitatively explore the habitability of Ganymede, Europa, and Titan. The next pressing challenge in ocean world astrobiology is biosignature identification and detection with these missions' successors[1,8,9].

The National Academies' Origins, Worlds and Life Decadal strategy for 2023-2034[8] emphasises that assessing biosignature suitability requires a comprehensive approach which includes understanding abiotic baselines, characterising possible context-appropriate biotic and abiotic processes, and evaluating proposed biosignature robustness by always considering alternative lines of evidence. In light of these recommendations, recognising and accounting for a target's abiotic baseline has emerged as a powerful component in the search for life[10]. Several candidate biosignatures and life detection strategies have been proposed for icy moons, including mass distributions in organic matter[11], chirality[12,13], direct cell observation[14], chemical species relevant to metabolism and early biology[15,16], and isotope fractionation[17,18]; however, each of these can be obfuscated by abiotic processes. To maximise the confidence in any hypothesis suggesting biological activity, and to minimise the likelihood of false positive and negative conclusions, it is essential to carefully evaluate the full spectrum of both abiotic and biotic potential contributions to observed (bio)geochemical signatures[8] (**Fig. 1**). One way to achieve this is by building an abiotic baseline model based on existing knowledge, then determining whether hypothetical biotic contributions can be inferred above that baseline.

This article presents an adaptable framework that emphasises understanding the abiotic baseline before considering biology. The framework is designed for quantifying the abiotic baseline of any ocean world with respect to any arbitrary biosignature. Briefly, following from box models of biogeochemical cycling on Earth[19], the framework has a modular structure that allows for specific processes and features to be added or omitted dependent on the target world. To demonstrate its utility, this article explores an example on carbon isotope and amino acid enantiomer cycling on Saturn's moon Enceladus.

Enceladus is thought to have a vast ocean of liquid water situated between its icy shell and rocky core (**Fig 2**). Plumes erupting from Enceladus' south polar terrain include icy grains containing salts, silica, and a variety of organic compounds[20–24] as well as gases including $CO_2$, $H_2$, $CH_4$, $NH_3$ and $H_2O$[6,25]. The co-occurrence of $H_2$, $CO_2$, and $CH_4$ is astrobiologically significant, as these chemical species are the substrates and products of hydrogenotrophic methanogenesis, an ancient metabolism on Earth. This raised the hypothesis that Enceladus could be habitable for organisms using this metabolism known as methanogens[3–6]. Although several additional metabolic pathways have been suggested for Enceladus[26–28], methanogenesis is thought to offer the highest potential for biomass due to its energy yield[26], and is unique in that all of its required reactants have been directly observed[6]. However, because $H_2$, $CO_2$ and $CH_4$ are produced and consumed by both biotic and abiotic processes, determining their origin requires a quantitative framework that accounts for all plausible non-biological pathways.

The Enceladus example establishes an empirical abiotic baseline of methane ($CH_4$) and its carbon isotopic relationship with carbon dioxide ($CO_2$) based on Enceladus' geologic and geochemical context, which is then compared against a hypothetical biotic contribution to that relationship. Racemization of chiral amino acids contributed by the same hypothetical biological community is also included and monitored through ocean transport. These potential biosignatures were selected because, as discussed above, $CH_4$, $CO_2$ and other organic material have been detected on Enceladus[6,20,21,25] and participate in the methanogenesis metabolism, and $CH_4$'s abiotic and biotic formation and evolution processes on Earth are well characterised (**Table 1**). Focusing on simple metabolic reactants and products is also beneficial because it potentially adds a readily deployable independent line of evidence for life detection[8] and complements methods designed to estimate biosphere size and productivity[4,26,29,30]. Furthermore, even if most of the $CH_4$ observed on Enceladus originates directly from methanogenic archaea, the anticipated associated productivity of its biosphere would likely be small[4,29]. Methanogen cell abundances may therefore be too low to reliably detect, should they even survive the journey from seafloor to surface intact[4,29]. While carbon isotopic uncertainty from existing observations on Enceladus is high (>150‰)[25], and enantiomeric excesses are unknown, the framework identifies the key gaps in mechanistic understanding that need to be filled to best interpret future observations and is extensible to other ocean worlds.

## Results

A description of the framework's modular design is included in the **Methods**. The Enceladus example is first systematically described in detail through the lens of carbon isotope cycling (visualised in **Fig. 2**), then amino acid chirality is explored as a second potential line of evidence which can be assessed within the same abiotic baseline framework.

### *Enceladus' abiotic baseline for carbon isotopes*

An abiotic baseline for carbon isotope cycling on ocean worlds must consider all carbon sources and sinks, all possible rate-limiting and carbon isotope selective steps during each process, and their corresponding kinetic isotope enrichment factors specific to the local physicochemical settings (*e.g.,* temperature, pressure, pH). Isotope enrichment factors $\epsilon$ are used to compute the effect of isotopically selective processes at the mechanistic level, resulting in a net isotope difference $\Delta\delta^{13}$C between chemical species and at different locations on Enceladus (**Fig. 2**). The calculated net isotope difference mimics hypothetical observations that are influenced by multiple processes, carbon sources or carbon sinks. Detailed definitions of isotope terminology are listed in **Extended Data Table 1.**

**Table 1** summarises possible abiotic sources of $CH_4$ on ocean worlds, following reviews of processes on Earth[31,32], as well as those hypothesised for other worlds (*e.g.,* Enceladus[33] and dwarf planets[34]). A detailed description of these processes is provided in the **Methods**. The most probable and quantifiable non-primordial $CH_4$ sources on Enceladus at present are Fischer-Tropsch-Type (FTT, including Sabatier) reactions and thermal alteration of organic material, because others may be reasonably discounted based on likely *in situ* conditions (*e.g.,* those that necessitate temperatures >250°C, Enceladus' approximate hydrothermal temperature limit[35]; see **Methods**).**Extended Data Fig. 1c,d** compiles $\epsilon$ values for $CH_4$ production from these sources observed in laboratory and field studies[36–43]. For FTT/Sabatier $CH_4$ formation, the catalyst, carbon source, openness of the system, extent of reaction, and temperature affect the reported enrichment factors which, in this compilation, vary between -49.5‰ and -15‰, or -49.5‰ and -30‰ isolating T<250°C.

Thermal alteration of organic matter is another source of $CH_4$ on Earth, but the source organic material (kerogens) is overwhelmingly biological in origin. On icy worlds like Enceladus, $CH_4$ could also alternatively derive from thermal alteration of abiotic organic material[33]. Laboratory experiments[43] measuring such $CH_4$ produced from Murchison meteorite material suggest $^{13}C$ enrichment is very small (smaller than -5‰; **Extended Data Fig. 1d**), but this may be due to limited experimental data mostly focusing on high temperatures and short time periods (**Methods**). In contrast, on Earth, the net isotopic difference between kerogens in source rocks and associated $CH_4$ is between -24‰ and +5‰, depending on maturity, kerogen type, and local physicochemical settings[41]. To be maximally conservative in the abiotic baseline, this range is adopted as $\epsilon_{OM/CH4}^{TA,KIE}$ for $CH_4$ originating from thermal alteration of any precursor organic matter. In addition to FTT/Sabatier reactions and thermal alteration, a fraction of the abiotic $CH_4$ in Enceladus' ocean may be primordial[33] and mixing with any $CH_4$ subsequently produced *in situ*. For example, carbon observed on comets has $\delta^{13}$C values between $\pm{\sim}225‰$[44], and can vary temporally through alteration processes and escape[45].

While Enceladus' bulk ocean temperature is close to 0°C[2], temperature increases in proximity to proposed habitats, for example as ocean water mixes with hydrothermal fluid or infiltrates into the core[3], will affect the source $\delta^{13}C_{CO2}$ for any microbial activity or FTT/Sabatier reactions via the temperature-dependent carbonate speciation[3] and carbon isotope enrichment factors[46]. Carbonate species in Enceladus' ocean are likely fully dissolved[4,6,47,48], so the primary isotope exchange in seawater is between aqueous carbonate species. In this interaction, temperature–and not other geochemical uncertainties such as pH–primarily controls the net isotope difference in $CO_2$ throughout Enceladus' ocean (**Extended Data Fig. 2**). For example, for $\delta^{13}C_{CO2} = 60‰$ in Enceladus' ocean, the modelled temperature-driven net isotope difference is <8‰ between 0-100°C for bulk ocean pH 7-11 (**Extended Data Fig. 2**), and is relatively insensitive to different ocean $\delta^{13}C_{CO2}$ values.

The isotopic composition of $CO_2$ and $CH_4$ may also change during transit between the seafloor (beginning with an ocean plume) and the space plume. This includes vertical transport through tens of kilometres of ocean and exsolution at or near the ocean-ice shell interface. As Enceladus' ocean may be vertically stratified[49–52], three distinct layers are modelled across four depths: $x_1$ (seafloor) $x_2$ (maximum height of ocean plume), $x_3$ (ocean-ice shell interface) and $x_4$ (space plume) (**Fig. 2**). Between $x_1$ and $x_2$, vertical transport is efficiently conserved via rotating convection (referred to as the ocean plume). Between $x_2$ and $x_3$, this efficient vertical transport halts, and becomes restricted to turbulent vertical diffusion[50] or molecular diffusion (**Methods**). Between $x_3$ and $x_4$, gases exsolve into the space plume.

While bulk fluid flow (*e.g.,* advective flow) is not isotopically selective, different isotopologues can have different molecular diffusion coefficients $D_{i,j}$, and an associated species-specific enrichment factor $\epsilon_i^{diff,KIE}$. If molecular diffusion contributes to vertical transport, the isotopic composition may therefore vary spatially and temporally throughout the ocean column. In the ocean plume ($x_1$–$x_2$), Péclet numbers based on literature ascent velocities[12,53] and a minimum endmember based on space plume escape rates are all larger than 10,000, indicating that advection contributes significantly more to vertical transport than molecular diffusion in this region (**Extended Data Fig. 3**). In all tested Enceladus scenarios between $x_1$ and $x_2$, negligible net isotope differences owing to molecular diffusion are predicted (*i.e.,* $\Delta\delta^{13}C_{x1,CO2/x2,CO2} = \Delta\delta^{13}C_{x1,CH4/x2,CH4} = 0‰$; **Extended Data Fig. 4**). Notably, it is possible to have non-zero net isotope differences on alternative worlds where ascent velocities are at least 100 times lower than this minimal estimate for Enceladus and lateral loss processes are present (**Extended Data Fig. 4**).

The domain $x_2$–$x_3$ considers a possible stratified layer in Enceladus' ocean where advective flow ceases at $x_2$ and vertical transport slows to between an upper limit defined by turbulent vertical diffusion[50] and a lower limit defined by molecular diffusion (**Methods**). Close to the upper limit, the timescale of vertical transport is comparable to other estimates (greater than tens to hundreds of years[54], **Extended Data Fig. 5**, **6**), and a negligible $\Delta\delta^{13}\mathrm{C}_{x2/x3}$ is predicted for both $CO_2$ and $CH_4$. If molecular diffusion is the only mechanism of vertical transport in this domain, then both the bulk chemical composition and net isotope difference between $x_2$ and $x_3$ can be non-negligible depending on the time elapsed since layer formation and the layer depth (**Extended Data Fig. 5**). After bulk volatile concentrations at $x_3$ exceed ~10% of their concentration at $x_2$, the net isotope difference between $x_2$ and $x_3$ is $< 4‰$ for $CH_4$, and $< 2‰$ for $CO_2$ (**Extended Data Fig. 6**).

However, importantly, this molecular diffusion-only scenario is an extreme endmember, and a kilometres-thick layer could require millions to billions of years for vertical transport. This scenario is inconsistent with observations unless all observed $CH_4$ is primordial, or Enceladus has had a turbulent history and space plume observations reflect Enceladus' geological past, not necessarily its present.

For the transition between $x_3$ and $x_4$, the maximum $\Delta\delta^{13}\mathrm{C}_{x3/x4}$ is -1.3‰ for $CO_2$ and -0.7‰ for $CH_4$. This corresponds to the aqueous-gaseous phase transition only (**Methods, Fig. *2***). As transport from $x_1$ to $x_3$ described above considered dissolved $CO_2$ and $CH_4$, if gas bubbles form below $x_3$–which is possible but not certain in Enceladus' geochemical parameter space[4,55]–the isotopic composition at that depth would be the source for space plume gases.

**Fig. 2** summarises the framework and modelling results described above. Overall, systematically building a carbon isotope-focused abiotic baseline model for ocean worlds and applying it to Enceladus has identified that the largest contributor to uncertainty on this satellite is the source of abiotic $CH_4$, not other subsequent processing. Carbonate speciation and migrative processes from seafloor to space plume have a calculable, mostly marginal effect on $CO_2$ and $CH_4$ isotopic composition except in extreme endmembers. This interpretation is possible for the Enceladus example because all carbon species and their isotopic compositions were monitored from original source through to the observable region. Knowing the $\delta^{13}$C values of initial carbon sources available for (bio)geochemical reactions (in this model, dissolved inorganic carbon in the ocean and refractory organics in the core) is critical on any world for interpreting an observed $\delta^{13}$C value and identifying possible processes contributing to any calculated net isotope difference. Additionally, the source $\delta^{13}$C, and isotope fractionation factor, will vary significantly dependent on the target body and its history and require laboratory studies specifically tailored to similar conditions. On ocean worlds, the original core composition, pressure, and temperature, and how they have changed through time will affect both the bulk chemical composition and the net isotope difference between reservoirs such as subsurface oceans and fluids circulating in the core.

## *Possible characteristics of biological $CH_4$ on Enceladus*

Empirical studies on microbial $CH_4$ have identified $\epsilon^{bio,KIE}_{CO2/CH4}$ values for varying temperatures, pressures, pH values, microbial strains, and compositions[56–59] (**Extended Data Fig. 1b**). Enrichment factors are often more negative in mesophilic methanogens than for thermophiles and hyperthermophiles[56,57]. Additionally, enrichment factors across different strains vary with the affinity of the net methanogenesis reaction, peaking at approximately 50 kJ/mol then decreasing at both lower and higher affinities (ref [56], **Extended Data Fig. 7)**. In this work, the enrichment factor associated with microbial methanogenesis in an Enceladus-like habitat $\epsilon^{bio,KIE}_{CO2/CH4}$ is estimated using published habitat chemical speciation models[3,4,6,47] and the enrichment factor as a function of [$CO_2$] and [$H_2$] at 60°C[56] (**Methods**). When compared to the variety in enrichment factors that have been measured empirically, the expected $\epsilon^{bio,KIE}_{CO2/CH4}$ associated with biological hydrogenotrophic methanogenesis on Enceladus would be small, between -27‰ and -4‰ if the ocean has a pH value of 8, and -27‰ to -7‰ if the ocean pH is 9 (**Extended Data Fig. 1a**). This is caused by the comparatively high [$H_2$] in Enceladus habitats[3,4] compared to empirical settings that support a larger enrichment factor (up to -100‰[56], **Extended Data Fig. 7**). The Enceladus carbonate speciation model assumes that the [$H_2$]:[$CO_2$] ratio observed in the space plume is approximately preserved from the fluids emerging from the seafloor habitat[3,6]. Isotope enrichment by microbial methanogenesis could therefore be stronger if the [$H_2$]:[$CO_2$] ratio is 1-2 orders of magnitude smaller in the ocean or habitat than in the space plume. This could be enabled by preferential volatilisation of $H_2$ over $CO_2$ and $CH_4$; while, Enceladus ocean models containing dissolved gases published to date have not yet predicted this required change in $[H_2]$:$[CO_2]$[6,48,55,60], further studies are required to understand and refine the exact processes determining the plume's composition.

## *Uncertainty in Enceladus' carbon isotope abiotic baseline obscures a hypothetical biosphere*

**Fig. 3a-d** shows an example of how the framework can interpret hypothetical observations of $\delta^{13}C_{CH4}$ and $\delta^{13}C_{CO2}$ in Enceladus' space plume. Beginning with a hypothetical observation of $\delta^{13}C_{CO2}$ at 60‰, first the $\delta^{13}C$ of the carbon source for $CH_4$ is calculated using the ocean transit and speciation models. Next, this is converted into endmember values of $\delta^{13}C_{CH4}$ owing to each formation process (FTT/Sabatier, thermal alteration, and microbial). Finally, this is propagated back into the space plume resulting in windows of contributions to 'observed' $\delta^{13}C_{CH4}$ owing to each formation process. In classical diagnoses of biotic contributions to materials, mixing lines (*e.g.,* **Fig. 1b**) can be drawn between the 100% biotic and 100% abiotic endmembers, but importantly, in order to be effective, these require sufficient distinction between the two endmember values. On Enceladus, the modelled ranges in $\delta^{13}C$ signatures of 100% biotic $CH_4$ and 100% abiotic $CH_4$ against bulk ocean $CO_2$ substantially overlap. For the 60‰ example, endmembers in **Fig. 3a-c** in order are $[12‰ < \delta^{13}C_{\mathrm{x4,CH4}}^{FTT,abio} < 58‰]$; $[37‰ < \delta^{13}C_{\mathrm{x4,CH4}}^{TA,abio} < 79‰]$; and $[35‰ < \delta^{13}C_{\mathrm{x4,CH4}}^{bio} < 63‰]$, where superscripts denote $CH_4$ formation processes: *abio* signalling abiotic, and *bio* signalling biotic. **Fig. 3d** shows an additional hypothetical contribution where $CO_2$ incorporated into biomass produces $CH_4$ through thermal alteration, which ranges between $11‰ < \delta^{13}C_{\mathrm{x4,CH4}}^{TA,bio} < 69‰$.

Specifically, $CH_4$ production via FTT/Sabatier reactions on Enceladus may produce $\delta^{13}C_{CH4}$ similar to, or even more depleted than, microbial $CH_4$. Likewise, the thermal alteration of abiotic organic material may produce $\delta^{13}C_{CH4}$ similar to, or more enriched than microbial $CH_4$, but only if the $\delta^{13}C$ of the source organic material is similar to the inorganic pool used for methanogenesis (for comparison, the $\delta^{13}C$ of insoluble organic matter in meteorites varies between approximately -5‰ to -35‰[61]). Cumulatively, these result in an abiotic baseline with a $\delta^{13}C_{CH4}$ range which fully encompasses that of any anticipated microbial $\delta^{13}C_{CH4}$ signature on Enceladus before even considering primordial $CH_4$. In other words, measuring just $\delta^{13}C_{CO2}$ and $\delta^{13}C_{CH4}$ cannot yet realistically be used to infer any potential biosphere on Enceladus due to a highly uncertain abiotic baseline coupled with a weak potential biotic contribution. It may, however, help constrain the processes behind abiotic $CH_4$ formation, if observations within ~10‰ are feasible. While this was not possible with the Cassini spacecraft (uncertainty in $\delta^{13}C$ in Enceladus' plumes is approximately $\pm150$‰[25]), future spacecraft such as Europa Clipper will have improved analytical accuracy and precision for $\delta^{13}C$ observations[62,63], but exact uncertainties will depend on flyby velocity and plume composition.

### *False negative life detection risk for amino acid chirality observations*

Life is selective in the chirality of its amino acids, utilising primarily their Levorotatory (*L*-)form[64,65]. Abiotic amino acids are typically close to racemic, containing approximately 50% in the *L*-form and 50% in the Dextrorotatory (*D*-)form (**Fig. 1a,c**). Over time, mixtures with enantiomeric excess can undergo racemization, yielding racemic mixtures (*i.e., L*-form$=50\%$). This process was included in the framework for mixtures of biotic and abiotic amino acids transiting through ocean worlds.

Two temperature-dependent racemization rate constants were used to capture uncertainty in the rate of this process[66,67] (**Methods**). In the fastest racemization endmember, even at 0°C bulk ocean transport timescales may allow for significant racemization beginning at 100 yr, and biotic amino acids would be racemic within 10,000 yr (**Fig. 3f**). This timeline is shorter if the fluids have circulated in the 60°C habitat (**Fig. 3g-h**), as racemization rate increases with temperature[66,67]. Therefore, even if rapid advection controls transport through most of the ocean, a stratified upper layer just 1 km thick where turbulent diffusion dominates would be sufficient for total racemization if concentrations at the top of the regime are 90% those at the bottom (**Extended Data Fig. 6**). Shallower layers are also sufficient if molecular diffusion dominates over turbulent diffusion, as the diffusion coefficient for amino acids are a similar order of magnitude to those of $CO_2$ and $CH_4$[68]. Additionally, as Enceladus may host significant abiotic organic chemistry[12,20,69], hypothetical biotic amino acids on Enceladus may mix with abiotic ones, both reducing the initial enantiomeric excess and decreasing the time required for the mixture to become racemic (**Extended Data Fig. 8**). In summary, at present, uncertainty in abiotic processes–specifically timescales of ocean transport, circulation in subseafloor hydrothermal systems, and the contribution of abiotic organic material–suggest that enantiomeric observations on Enceladus could, within abiotic uncertainty, lead to false negative life detection conclusions.

## Discussion

Detecting credible evidence of life beyond Earth depends on our ability to unambiguously distinguish biosignatures above an abiotic baseline.  Establishing a rigorous abiotic baseline is thus increasingly recognized as fundamental to interpreting future biosignature observations. This article presents a flexible, quantitative and customizable framework to evaluate the abiotic geochemical and geophysical context of any ocean world. To demonstrate its potential, we apply it to Enceladus, focusing on whether carbon isotope or chiral biosignatures could be identified above that baseline with future observations. The Enceladus examples shows that both biosignatures may be obscured by abiotic processes, suggesting ambiguous interpretations for any carbon isotope observations, and risking false negative conclusions from chiral observations. Crucially though, as the framework is designed systematically using process-level understanding, it is possible to identify which physical and chemical properties need targeted focus to reduce uncertainty in Enceladus' abiotic baseline, and in doing so potentially reduce the likelihood of ambiguous observations identified in this work. Additionally, by including multiple potential biosignature types within a unified process-based model, the framework can simultaneously handle multiple lines of evidence for life detection. This allows for expanding the scope of this example to other complimentary biosignatures in the future, such as for multi-element isotopic compositions (e.g., H, O, and S), carbon isotopes in organic material, preservation and direct observation of partial cells, or patterns in amino acid abundances. Evaluating multiple biosignatures within the same geophysical and geochemical framework will help identify the most effective suite of instruments required to avoid ambiguous results from a life detection mission to any ocean world.

The Enceladus example includes contributions from: ongoing abiotic $CH_4$ production from organic and inorganic carbon sources, chemical interaction between carbonate species, biological $CH_4$ production, transport through the seafloor and ocean, and mass transfer into Enceladus' space plume. Additional uncertainty in establishing Enceladus' abiotic baseline for $\delta^{13}C_{CH4}$ could arise from a potential contribution by primordial $CH_4$, or additional $\delta^{13}C$ differences between abiotic organic matter and inorganic carbon. Any mixing with primordial $CH_4$ adds uncertainty over and above that reported in **Fig. 3a-d**, and similarly, a difference in $\delta^{13}C$ between organic carbon and inorganic carbon will introduce a systematic shift in the $\delta^{13}C_{CH4}$ window shown in **Fig. 3b**. Ambiguity in Enceladus' carbon isotope abiotic baseline may be reduced by ruling out processes based on yield, for example, $CH_4$ from thermal alteration of abiotic organic matter might be minor if $CO_2$ is its overwhelmingly dominant product[43,70] (see also **Methods**).

Results from the Enceladus example show that the isotope enrichment factor between $CO_2$ and $CH_4$ owing to methanogenic archaea, if present, is likely smaller than typically considered in Earth-like environments (**Extended Data Fig. 1a**), though recent empirical data support this in reducing, carbon-limited settings[71]. Importantly, the biological enrichment factors and predicted net isotope differences presented here necessarily consider only the hypothesis that Enceladus is inhabited with Earth-like methanogens and do not consider prebiotic settings, early evolutionary states, diverse ecologies, or otherwise alternative life-forms which may not behave similarly to methanogens in Earth laboratories[72]. Microbial methanogenesis at lower temperatures than considered here is also possible (including at 0°C[73]), but isotopic and kinetic data is more limited for methanogens in these settings. At <45°C, similarly small values of $\epsilon^{bio,KIE}_{CO2/CH4}$ are anticipated at the large methanogenesis affinities expected in Enceladus' bulk ocean[3,56], but explicit dependence on $[H_2]$ and $[CO_2]$ is not available at these temperatures. As models continue to develop, this framework can incorporate contributions from different and more complex microbial communities under different environmental conditions owing to the flexibility of flux-based microbial kinetic modelling[74].

The Enceladus example demonstrates how a systematically defined abiotic baseline can not only obfuscate biosignatures, but also diagnose current uncertainties and unknowns in geophysical and geochemical properties which future missions and models must resolve. For example, Enceladus' ocean transport timescale is of minor importance for $\delta^{13}C_{CH4}$ and $\delta^{13}C_{CO2}$, but is the key abiotic process determining whether enantiomeric excesses might be preserved to the ocean-top. This approach is transferrable to other ocean worlds, where modules can be added, adjusted, or removed dependent on local context. Europa's larger and more saline ocean may support buoyant convection with much shorter transport timescales than Enceladus[54,75], potentially omitting the need for a slow, diffusive transport zone, but less is currently known about the potential for ocean-ice exchange and transport on Europa, so additional modules may be needed (i.e., between $x_3$-$x_4$, **Fig 2.**). Meanwhile, Titan may have a subsurface ocean bounded by layers of ice[76], precluding hydrothermal activity and fast ocean transport, but additionally hosts complex abiotic organic chemistry on its surface and in its atmosphere[77], and possibly in its interior[78]. This abiotic organic chemistry is important not only for defining Titan's abiotic baseline, but also as a possible carbon and energy source for life, provided it can be transported to the ocean[30]. The process-based nature of this framework and its focus on geophysical and geochemical variables and their uncertainties allow it to be flexibly applied and extended to other worlds.

The modular framework that incorporates multiple locations or layers can also help define targets and performance requirements for future life detection missions. For example, on Enceladus, there would be a higher risk of false negative results when measuring chirality by sampling the space plume or surface fallout than by an ocean probe that samples below the diffusive layer (*i.e.,* at $x_2$, **Fig. 2**). Quantitative information on internal heat dissipation, thermal gradients, densities, and layering would improve precision in Enceladus' abiotic baselines for both carbon isotopes and racemization, and could be achieved by better understanding tidal deformation (*e.g.,* resolving unconstrained Love numbers)[79]. Combining this with modelling constraints on the dynamics of freeze/thaw cycling in Enceladus' ice shell will improve understanding of vertical transport in the ocean[49–51,54,80,81]. More broadly, characterising Enceladus' bulk abiotic organic inventory–both primordial and any ongoing production–is also important for identifying abiotic $CH_4$, discriminating potential biological organics, and as another possible energy source for life[30].

Overall, this framework and its application to Enceladus reveals that contextual abiotic geochemistry and geophysics must be prioritized for all ocean world life detection strategies. Geochemical and geophysical processes enable sustained habitability[82], and the systematic, quantitative construction of an abiotic baseline framework in this work demonstrates that they are also critical for reducing ambiguity in biosignature observations. To ensure the most reliable path to life detection, mission objectives such as characterising planetary-scale temperature gradients, core lithology, rheology, and ocean density should therefore be considered a necessary supplement to direct biosignature observation.

# Methods

## *Structure for the abiotic baseline framework*

The abiotic baseline framework follows a modular structure (**Fig. 2**). The modules communicate by mass flux, allowing separate mechanisms to be modelled both individually and as a segment of the wider system. This technique is common in simple models of Earth's biogeochemical cycles[19,83] and a specific requirement for isotope cycling is that the modules are in a steady state condition[83]. Mass balances are constructed as:

$$J_{in} + \sum J_{inside} + J_{out} = 0 \quad (1)$$

for the bulk composition, and:

$$R_{in}J_{in} + \sum R_{inside}J_{inside} + R_{out}J_{out} = 0 \quad (2)$$

for the minor isotope. In Eq ( 1 ) and ( 2 ), $R$ is the fractional relationship between the heavy isotope and the light isotope taking part in a process (*e.g.,* $R = {}^{13}C / {}^{12}C$), and $J$ is a flux in mol/s, which is positive for input processes and negative for removal processes. It is assumed throughout isotopic analysis that the bulk composition properties approximate to those of $^{12}C$, as the C pool is mainly comprised of $^{12}C$ (*i.e.,* $^{13}C \ll {}^{12}C$). Values of $R$ were converted to $\delta^{13}C$ using the Vienna Pee Dee Belemnite standard. At steady state and under first-order chemical kinetics, equilibrium and kinetic isotope effects are mathematically equivalent, so isotope enrichment factors between carbon source $A$ and its produced $CH_4$ are utilised as $\epsilon_{A/CH4} \equiv \frac{R_A}{R_{CH4}} - 1$. The corresponding net isotope difference for $CH_4$ generated by this process is $\Delta\delta^{13}C_{A/CH4} = \delta^{13}C_A - \delta^{13}C_{CH4}$. Isotope definitions are summarised in **Extended Data Table 1**. The $\epsilon$ and computed $\Delta\delta^{13}$C values for individual modules of the Enceladus example are shown in **Fig. 2.**

## *Abiotic $CH_4$ sources on ocean worlds and their isotopic properties*

Determining the abiotic vs biotic origins of organic material on Earth, particularly $CH_4$, is challenging[32,84–87]. Following the assignments by Etiope & Sherwood Lollar (2013)[84], **Table 1** categorises $CH_4$ sources on ocean worlds with notes specific to Enceladus. This summary is necessarily more general than previous Earth-focused reviews[31,32,84,86,88] and includes important additional relevance to other ocean worlds. There are three categories of abiotic $CH_4$: primordial $CH_4$, $CH_4$ derived from inorganic carbon, and $CH_4$ derived from abiotic organic carbon.

Primordial $CH_4$ was either already present in-form in Enceladus' precursor material or was created during its formation and differentiation. Major categories of primordial $CH_4$ are:

1. *Cometary.* $CH_4$ and methanol make up the bulk of organic material in the ices of molecular clouds which formed the solar system bodies[89], so may have been trapped in clathrates in the planetesimals that constituted Saturn's feeding zone[90]. Future noble gas measurements (particularly Xe) may help determine how much primordial $CH_4$ could have been delivered in this way[90]. $CH_4$ is also present in comets and meteorites. It's abundance varies in cometary ices between ~0.1-1% (relative to $H_2O$) and is typically significantly lower than CO and $CO_2$[91]. The carbon isotope distribution in meteorites and comets is complex and may change through time, resulting in large variations between measurements[45]. For example, chondritic meteorites and comets can produce additional $CH_4$ in their interiors, where radiogenic heating allows an internal water phase, releasing other volatiles (e.g., CO, $H_2$, $CO_2$) – whose $\delta^{13}C$ appears to depend on their isotopic equilibrium at formation[92]. Slow FTT reactions may then proceed (see below) to generate $CH_4$ and longer chain hydrocarbons from the $CO_2$ and CO[92,93]. In some cases this isotopically light $CH_4$ may be lost, enriching the remaining carbonates and $CH_4$[e.g., 94]. The Murchison Meteorite has the following $\delta^{13}C$ values: 29.1‰, -31‰, 9.2‰ for $CO_2$, CO and $CH_4$, respectively[95]. These multiple origins and variation in the limited data available for cometary volatiles make it challenging to determine what the isotope signature of purely primordial $CH_4$ on Enceladus would be. Measurements of $\delta^{13}C_{CH4}$ on comets such as 67P (14±116‰[96]) or $CH_4$ ices on dwarf planets (e.g., Eris: 68±178‰, and Makemake: -110±267‰[34,97]), are comparable within uncertainty to the Enceladus bulk $\delta^{13}C$ estimate by Waite et al., (2009)[25].

2. *Refractory organic material.* $CH_4$ and other reduced hydrocarbons may be formed from organics in Enceladus' accretional material via thermal alteration (discussed below). Outer solar system bodies likely accreted from carbonaceous asteroids and comets[98], which contain a variety of complex refractory organics[61], so comets and carbonaceous chondrites are one possible analog for this starting material. Additionally, Saturn's largest moon Titan has a vast inventory of abiotic refractory organic material produced via photochemistry in its atmosphere[77]. If the photochemistry in Titan's atmosphere is a good analog for Enceladus' primordial refractory organics, then isotopic measurements of laboratory analogs of Titan haze could be an alternative analog for Enceladus $CH_4$ source material.
3. *Magmatic $CH_4$.* $CH_4$ is formed in Earth's mantle at very high temperature (>500°C) from carbides, and at lower temperatures through respeciation of C-O-H[84]. Due to Enceladus' small size, it is unlikely that temperatures in these range were ever present (Enceladus' maximum core temperature is ~700 K (427 °C)[99,100]).

To capture the spread of possible isotope enrichment factors for $CH_4$ production from inorganic carbon, values from a wide range of experiments under different phases and temperatures and with different catalysts and C sources were compiled together (**Extended Data Fig. 1**). These are briefly described below.

$H_2$ generation by serpentinization and subsequent reaction with inorganic carbon to $CH_4$ has been suggested as one potentially ongoing abiotic source of Enceladus' $CH_4$[5,7,33]. Similar processes are thought to generate abiotic $CH_4$ on Earth (**Table 1**), though the exact mechanisms, yields, and isotope fractionation observed in the laboratory varies significantly. Typically, FTT reactions are invoked, which take the following form (for $CH_4$):

$$3H_2 + CO \rightleftharpoons CH_4 + H_2O$$

A common FTT reaction is the Sabatier reaction:

$$4H_2 + CO_2 \rightleftharpoons CH_4 + 2H_2O$$

Aqueous FTT/Sabatier reactions have been suggested as one mechanism to generate $CH_4$ in Earth's hydrothermal systems, but these experimental studies often do not generate large yields compared to dry experiments performed in the gaseous phase[31]. Hypothetically, at high pressure but relatively low temperature conditions (*e.g.,* <150°C), enhanced $H_2$ and $CO_2$ solubility might change this observation, but is yet to be empirically tested. Aqueous FTT proceeds via CO and/or HCOOH[40] through intermediate methanol and/or methane-thiols[101]. At hydrothermal temperatures, these intermediate species arise from the water-gas-shift reaction:

$$CO_2(aq) + H_2(aq) \rightleftharpoons HCOOH(aq) \rightleftharpoons CO(aq) + H_2O(l)$$

FTT/Sabatier reactions often require catalysts to overcome their significant kinetic barriers (e.g., Ni-Fe, Fe, Co[36–40,102]). While uncatalysed redox pathways to $CH_4$ have been reported when T>150°C[103] (including over multi-year timescales at >300°C at comparable rates to mineral-catalysed experiments[101]), their $^{13}C$ isotope enrichment factors have not yet been measured. Even when mineral catalysts are present, hydrothermal conversion of aqueous carbonates to $CH_4$ can be kinetically sluggish, owing to poor access to mineral surfaces, relatively stable intermediate species, or $H_2$ and $CO_2$ solubility[32,103], but yields can improve when some component of the $H_2$ and $CO_2$ is in the vapor phase[32]. Experiments aiming to determine whether low temperature FTT/Sabatier reactions from ultramafic rocks occur on laboratory timescales have been inconclusive and affected by contamination[104], but abiotic methane has been inferred in low temperature, highly reducing field sites on Earth[84], and formed at room temperature in the laboratory from gaseous $H_2$ and $CO_2$ with a Ru catalyst[105]. In some cases, alternative sources of $CH_4$ have been identified for the uncatalysed experiments (*e.g.,* fluid inclusions in olivine[106]).

Conditions in Enceladus' interior may support FTT/Sabatier reactions[33]. Metal catalysts, such as Ni and Fe, are likely present in Enceladus' core because they are common on meteorites, hydrothermal temperatures are predicted to be >90-200°C[35,107], and thermodynamic models have shown that pressure near Enceladus' seafloor may allow for vapor-phase $H_2$ and $CO_2$[4]. While these conditions support the possibility for FTT/Sabatier reactions, it has been suggested that catalytic activity on Enceladus may have been lost on Enceladus over geological timescales[107]. Even if that is the case however, it does not discount FTT/Sabatier reactions as a source or methane from Enceladus' past, nor does it discount slower, non-catalysed $CH_4$ production. The hydrothermal temperatures predicted on Enceladus do however likely preclude the other abiotic sources listed in **Table 1**.

Some $CH_4$ in active hydrothermal systems on Earth may instead originate from magmatic $CO_2$ trapped in fluid inclusions with long residence times and therefore be independent of ongoing circulating serpentinising fluids[106,108], thus mitigating the kinetic problems outlined above. In this scenario, magmatic $CO_2$-rich fluids are trapped in fluid inclusions, and the water is slowly removed through serpentinization and hydrated into the rock, leaving gas phase $H_2$ and $CO_2$ to react slowly and generate $CH_4$[108,109]. Isotopic evidence supports this for numerous hydrothermal fields on Earth[108–110], though the isotopic signature of the source carbonates depends on the specific geochemical setting (*i.e.,* magmatic versus ancient seawater versus modern seawater).

Another category of abiotic $CH_4$ formation is through thermal alteration of organic carbon. This type of thermal alteration on Earth is often referred to as a biogenic process, because the source organic matter is overwhelmingly fixed by biology. However, on ocean worlds that may be scarcely inhabited or uninhabited, do not support photosynthesis, and/or have plentiful abiotic organic matter (as Enceladus does[21]), thermal alteration of abiotic organic material could be an important $CH_4$ source[33]. Pyrolysis experiments on insoluble organic material (IOM) in the Murchison meteorite demonstrate that $CH_4$ can be produced from the thermal degradation of abiotic organic matter[43]. However, as with experiments on converting inorganic carbon to $CH_4$, it is challenging to compare laboratory-timescale processes with phenomena that occur over longer timescales in nature. Okumura and Mimura (2011)[43] measured very small isotope enrichment factors between Murchison IOM and the $CH_4$ evolved (~-5‰; **Extended Data Fig. 1**), and $CH_4$ only evolved above 450°C, but the experiments were performed in 10 minute steps (and notably that isotope enrichment was larger than for a control kerogen). Recent experiments by Miller et al., (2025)[70] demonstrated $CH_4$ and $CO_2$ production from Murchison IOM down to 250°C, with a possible but currently unquantified increase in $\delta^{13}C_{CO2}$ relative to $\delta^{13}C_{CH4}$. Additionally, $CO_2$ was the dominant product (>99%) at reaction temperatures <500°C[70], so $CO_2$ could dominate the organic decomposition products on Enceladus and the $CH_4$ contribution could be small.

On Earth, the net isotopic difference between kerogens in source rocks and natural $CH_4$ evolved is between -24‰ and +5‰, dependent on maturity, kerogen type, and local physicochemical settings[41]. However, this is an *in situ*, post-production comparison, so the remaining kerogens will likely have a transient bulk $\delta^{13}C$ signature different from the original $CH_4$ source. Typical isotope enrichment factors between source kerogens and $CH_4$ measured in the laboratory are -10‰, varying by ±5‰ with temperature[42,111]. $CH_4$ production by thermal alteration of kerogens in nature begins at approximately ~50°C and peaks at ~150°C[41].

## *Carbonate speciation and isotope model*

The net isotope difference in $\delta^{13}C$ resulting from carbonate speciation is calculated using a carbon isotope mass balance model:

$$[DIC]R_{DIC} = [CO2]_T R_{CO2,T} + [HCO3]_T R_{HCO3,T} + [CO3]_T R_{CO3,T} \quad (3)$$

$$[DIC]R_{DIC} = [CO2]_T R_{CO2,T} + [HCO3]_T (1 - \epsilon_{HCO3/CO2}) R_{CO2,T} + [CO3]_T (1 - \epsilon_{CO3/CO2}) R_{CO2,T} \quad (4),$$

where $\epsilon_{\nu/CO2}(T)$ denotes an equilibrium isotope enrichment factor between carbonate species $\nu$ and $CO_2$ at temperature $T$ [K][46]. Values of $\delta^{13}C$ for $HCO_3^-$, $CO_3^{2-}$ and DIC (total dissolved inorganic carbon) are calculated at 0°C using Eq. ( 4 ) using published estimates for their concentration in Enceladus' bulk ocean[4], temperature-dependent isotope enrichment factors[46], and an input $\delta^{13}C_{CO2}$. Following the carbonate speciation procedure outlined in Higgins et al., (2024)[4,6,47], $\delta^{13}C$ for all carbonate species at elevated temperature are estimated by setting [DIC] as constant and allowing isotopes to equilibrate. Kinetic isotope effects may be important depending on the timescale over which seawater is heated, but as this is currently unconstrained for Enceladus it is assumed that isotope equilibrium is reached rapidly amongst dissolved carbonates, consistent with recent studies on the bulk composition[3,4].

**Extended Data Fig. 2** shows the change in $\delta^{13}C$ between the carbonate species with temperature and across Enceladus' concentration range of [DIC] (0.01 to 0.1 mol kg$^{-1}$), and accounts for varying [$Cl^-$] in the ocean model (0.05 to 0.2 mol kg$^{-1}$)[22,47].

### *Biological isotope model*

The mass balance for biological $CH_4$ production in an Enceladus habitat follows the Valentine et al. (2004) summary of a flow-through reactor at steady state[57]:

$$R_{\mathrm{CO2}}^{in} J_{\mathrm{CO2}}^{in} + \sum R_{\mathrm{CO2}}^{bio} J_{\mathrm{CO2}}^{bio} + R_{\mathrm{CO2}}^{out} J_{\mathrm{CO2}}^{out} = 0 \qquad (5)$$

$$R_{\mathrm{CH4}}^{in} J_{\mathrm{CH4}}^{in} + R_{\mathrm{CH4}}^{bio} J_{\mathrm{CH4}}^{bio} + R_{\mathrm{CH4}}^{out} J_{\mathrm{CH4}}^{out} = 0 \qquad (6).$$

Here, internal processes of $CO_2$ uptake $\sum J_{\mathrm{CO2}}^{bio}$ include the net metabolic pathway (*i.e.*, conversion to $CH_4$) and biomass production. The enrichment factor for microbial methanogenesis is designated $\epsilon_{\mathrm{CO2/CH4}}^{bio} = \frac{R_{\mathrm{CO2}}^{in}}{R_{\mathrm{CH4}}^{bio}} - 1$, and in the 100% biotic $CH_4$ production endmember case, $J_{\mathrm{CH4}}^{in} = 0$ and $R_{\mathrm{CH4}}^{out} = R_{\mathrm{CH4}}^{bio} = \frac{R_{\mathrm{CO2}}^{in}}{1+\epsilon_{\mathrm{CO2/CH4}}^{bio}}$.

To estimate the net isotope difference between $CO_2$ and $CH_4$ in a hypothetical Enceladus seafloor habitat, $\epsilon_{\mathrm{CO2/CH4}}^{bio}$ values reported from recent empirical studies[56] are matched to previously determined Enceladus-like habitat conditions[4]. A habitat temperature of 60°C is used because Gropp et al. (2022)[56] provide empirically determined isotope enrichment factors across [$CO_2$] and methanogenesis affinity values that overlap with inferred Enceladus-like conditions[3] at this temperature. Data from Gropp et al. (2022)[56], spanning $10^{-4}$ to$10^{-1}$ mol kg$^{-1}$ [$CO_2$] and affinities between 0 and 200 kJ/mol $CO_2$ are estimated using Clough-Tocher interpolation[112] and compared to Enceladus-like habitats from Higgins et al., (2024)[4]. This is visualised in **Extended Data Fig. 7**. Enrichment factors are calculated across the [$CO_2$] and [$H_2$] parameter space at pH 8 and 9 described in Higgins et al., (2024)[4], and the output distribution is shown as a boxplot for analysis and visualisation in **Extended Data Fig. 1**.

## *Fast, rotating ocean transport model; $x_1 \rightarrow x_2$*

Transport through Enceladus' hydrosphere is divided into three distinct layers at different depths $x_n$ (**Fig. 2**). Fast, rotating ocean transport (between $x_1$–$x_2$) is modelled using the one-dimensional advection-diffusion equation with an additional loss term:

$$\frac{\partial C_{i,j}}{\partial t} = v\frac{\partial C_{i,j}}{\partial x} + D_{i,j}\frac{\partial^2 C_{i,j}}{\partial x^2} - j_{loss,i.j}(x) \quad (7),$$

where $C_{i,j}(x,t)$ [mol dm$^{-3}$] is the concentration of isotopologue *j* for chemical species *i*, $v(x)$ is the bulk fluid's advective velocity [dm s$^{-1}$], $D_{i,j}$ is the molecular diffusion coefficient [dm$^2$ s$^{-1}$], *t* is time [s], *x* [dm] is the vertical distance from the seafloor, and $j_{loss,i,j}$ [mol dm$^{-3}$ s$^{-1}$] is a mass transfer coefficient for physical loss mechanisms during this transit (*e.g.,* any lateral transport that recycles into the ocean and does not directly contribute to observations) and hence can vary with *x*. Because the light isotopologue $^{12}C$ is substantially more abundant than $^{13}C$, $D_{i,12}$ for $CO_2$ and $CH_4$ is approximately equal to the molecular diffusion coefficient for the bulk molecular pool, which is well defined across salinity and temperature ranges[113,114]. Literature isotope enrichment factors[115–118] are used to calculate $D_{i,13}$ via the relation: $\epsilon_i^{diff,KIE} = \left(\frac{D_{i,13}}{D_{i,12}} - 1\right)$, and the most negative values of $\epsilon_{CO2}^{diff,KIE}$ are used here to estimate the maximum net isotope difference ($\epsilon_{CO2}^{diff,KIE} = -0.9‰$[116], and $\epsilon_{CH4}^{diff,KIE} = -3.3‰$)[117]. At steady state, the $j_{loss,i,j}$ term is required for any net isotope difference to be present between $x_1$ and $x_2$, owing to the conservation of mass flux at these module boundaries. A net isotope difference may be present when the $j_{loss,i,j}$ term is included because different isotopologues have different diffusion coefficients, and hence have different compositional depth profiles. Therefore, even if the removal process itself is not isotopically selective (*e.g.,* due to convection or other lateral removal processes), the removed fluid may have a different isotopic composition than is present at the top of this region ($x_2$) or the seafloor ($x_1$).

The Péclet number $P_e$ is a useful measure to compare the relative importance of diffusion versus advection and takes the form:

$$P_e = \frac{v(x)L_k}{D_{i,j}} \quad (8),$$

where $L_k$ is the length scale (in this case, the vertical distance between $x_1$ and $x_2$, denoted $L_1$). When $P_e \gg 1$, advection dominates fluid transport and when $P_e \ll 1$, diffusion dominates fluid transport. Setting the advective velocity at $x_2$ as $v_{x2} = \frac{r_w}{\rho A}$, where $r_w$ [kg $s^{-1}$] is the total mass of fluid moving through the column, $\rho$ [kg $dm^{-3}$] is the density of the fluid, and $A$ [$dm^2$] is the area through which it is moving, allows $P_e$ to be estimated for different endmembers of length scale and fluid flow rate (**Extended Data Fig. 3**). In all endmember conditions considered[12,53,119], including a conservative maximum coupled to a wide ocean column[119] and flow rates equivalent to Enceladus' space plume jet (100-1000 $kg_{H20}$ per second[120–123]), $P_e \gg 1$ suggesting an advection-dominated regime. This preliminary calculation informs the modelling procedure outlined below.

The significance of $j_{loss,i,j}(x)$ and $P_e$ on net isotope difference is modelled via a sensitivity analysis of different possible ocean configurations ($P_e$, *v(x)* and *A* as shown in **Extended Data Fig. 3**). In each configuration, Eq. ( 7 ) is solved numerically using the central difference method[124], using boundary conditions equivalent to $J_{in}$ and $J_{out}$ at $x_1$ and $x_2$ respectively. A steady state concentration profile in $x$ is calculated assuming advective-dominant fluid flow:

$$C(x) = \frac{J_{in}(x) + \int_{x1}^{x2} J_{loss}(x)dx}{v(x)A(x)} \quad (9).$$

In all simulations, the net isotope difference between $x_1$ and $x_2$ for both $CO_2$ and $CH_4$ are close to 0‰, owing to the large $P_e$ in this region. However, in some of the parameter space, solving Eq. ( 7 ) when $J_{loss} > 0.7J_{in}$ becomes numerically unstable due to very low concentrations in loss regions in the column. As these are the locations where any net isotope difference could be observable, to be maximally conservative, another algorithm was developed which solves for only the diffusive part of Eq. ( 7 ). In this scheme, any diffusive effect on the concentration profile in a time step *dt* is used to update $j_{loss}$ (*i.e.,* the concentration profile changes, but total fluid removal remains constant, changing the molar removal rate of species *i* and isotopologue *j*). To achieve mass balance, $J_{in}$ or $J_{out}$ is updated dependent on which of these fluxes are known, and an updated advective-dominant concentration profile is calculated (Eq. ( 9 )). The key control on $\Delta\delta^{13}$C in this conservative implementation is the simulation timestep *dt*. Von Neumann stability analysis[124] on Eq. ( 7 ) suggests a maximal stable *dt* of: $dt_{def} \leq \frac{dx^2}{v_{max}dx+2D_{i,j}}$, and the corresponding change in $\delta^{13}$C for $CO_2$ and $CH_4$ across Enceladus-like conditions is shown in **Extended Data Fig. 4**. In all Enceladus cases modelled, $\Delta\delta^{13}$C is still ~0‰, but extending the parameter space to hypothetical smaller Péclet numbers predicted some non-zero net isotope differences. This effect may therefore be important in the carbon isotope cycling of other, less active ocean worlds (*e.g.,* if ocean plume flow rates on Europa are significantly slower than for Enceladus).

### *Slow vertical transport through stratified ocean layer; $x_2 \rightarrow x_3$*

Between $x_2$ and $x_3$, bulk vertical fluid transport may become difficult[50,54]. An upper limit to transport has been suggested via turbulent diffusion (with a coefficient $D_t$~$10^{-4}$ $m^2$ $s^{-1}$)[50,54], but there is currently no constrained lower bound. Therefore, one candidate for the slowest form of vertical transport in this case could be molecular diffusion. Transport timescales and isotope-specific concentration depth profiles between $x_2$ and $x_3$ are calculated using a non-dimensionalised version of Fick's Second Law:

$$\frac{\partial \widetilde{C_{i,j}}(\tilde{x},\tilde{t})}{\partial \tilde{t}} = \frac{D_{i,j}}{(L_2)^2} t_{ref} \frac{\partial^2 \widetilde{C_{i,j}}(\tilde{x},\tilde{t})}{\partial \tilde{t}^2} \quad (10),$$

where $C_{i,j} = \widetilde{C_{i,j}} \times C_{i,j,ref}$; $t = \tilde{t} \times t_{ref}$; and $x = \tilde{x} \times L_2$. $C_{i,j,ref}$, $t_{ref}$, and $L_2$ are reference concentrations [mol $dm^{-3}$], timescales [s], and length scales ($L_2 = x_3 - x_2$) [dm] respectively. Setting the boundary condition $\widetilde{C_{i,j}} = 1$ at $\tilde{x} = 0$ has a well-defined solution[125]:

$$\widetilde{C_{i,j}}(\tilde{x},\tilde{t}) = \mathrm{erfc}\left[\frac{\tilde{x}}{2} \times \sqrt{\frac{(L_2)^2}{D_{i,j}\tilde{t}t_{ref}}}\right] \quad (11),$$

allowing $\widetilde{C_{i,j}}$ to be solved at any location in the domain at any $\tilde{t} > 0$. Furthermore, using the above boundary condition means that isotope ratios between $x_3$ (i.e., $\tilde{x} = 1$) and $x_2$ (*i.e.,* $\tilde{x} = 0$) can be directly calculated at any $\tilde{t} > 0$ using the nondimensionalised concentration at $x_3$: $\frac{R_{i,x3}}{R_{i,x2}} = \frac{\widetilde{C_{i,13}}(\tilde{x}=1,\tilde{t})}{\widetilde{C_{i,12}}(\tilde{x}=1,\tilde{t})}$. This is used to estimate the net isotope difference between $x_2$ and $x_3$ via the expression $\Delta\delta^{13}C_{i,x3/x2}[‰] = 1000 \times \left(\frac{R_{i,x3}}{R_{i,x2}} - 1\right)$, which is a reasonable approximation for $\delta^{13}C$ values close to zero.

To estimate the difference in bulk concentration at the top and bottom of the diffusive region, Equation ( 11 ) is solved for the $^{12}C$ isotopologues as a function of time and the distance $L_2$ in two cases: firstly, for $CO_2$ and $CH_4$ where molecular diffusion is the only method of vertical transport; and secondly, when turbulent diffusion contributes to vertical transport. This is achieved by varying the diffusion coefficient between $D_{i,12}$ and $D_t$+$D_{i,12}$. In the latter case, the curves for $CH_4$ and $CO_2$ are indistinguishable (**Extended Data Fig. 5a**). This is then repeated for the $^{13}C$ isotopologues and used to calculate $\Delta\delta^{13}C_{i,x2/x3}$ as described above (**Extended Data Fig. 5b**). **Extended Data Fig. 6** further dissects this by solving Equation ( 11 ) for $t$ when $x_3$ (*i.e.,* $\widetilde{x} = 1$) reaches certain concentrations, specifically 90% the value at $x_2$, and 10% the value at $x_2$. This shows the time required for each transport mechanism to deliver material to the top of this region.

## *Ocean-ice-space transition; $x_3 \rightarrow x_4$*

Mass balance for the transition between aqueous species at the ocean-top $x_3$ and gases in the space plume $x_4$ is calculated using equation ( 2 ), where the "loss" term encapsulates any fluxes that recirculate into the ocean, are incorporated into the ice sheet, or condense on the ice before escaping into its space plume. Several candidate mechanisms exist to explain how Enceladus' ocean connects to the space plume and the chemical fractionation that may occur on transit[6,21,48,55,126]. In cases where there is total mass transfer[6,47] or turbulent escape (*e.g.,* bubble scrubbing[126]), and any recirculation back into the ocean is not isotopically selective, no net isotope differences should be expected between $x_3$ and $x_4$. In other mechanisms, such as equilibrium or kinetic exsolution, there will be a net isotope difference in the gases that remain dissolved in the ocean and those that exsolve into the space plume (or equivalently, exsolve during ocean transit, which may be possible in shallow regions of Enceladus' ocean[4]). For $CO_2$, both kinetic and equilibrium enrichment factors between aqueous and gaseous phase are approximately -1‰[127], and for $CH_4$ they are between $\pm 1‰$[128,129].

Enceladus' ocean also may not be directly connected to the space plume, and recent work demonstrated that the observed escaping flux can be sustained via slushy melt from strike-slip motion in the ice sheet[130,131]. In this scenario, the observed plume gases must have been trapped in the ice then exsolve as the ice melts. To include this as a module in the framework would require including the incorporation of $CO_2$ and $CH_4$ into and out of the ice.

### *Amino acid racemization model*

Chiral molecules exhibit a specific type of isomerism. Their two isomers, known as *D*-form and *L*-form enantiomers, cannot be superimposed on their mirror image by any combination of rotations and translations. Enantiomeric excess in amino acids has been proposed as a biosignature, because abiotic amino acids are typically racemic (*i.e.,* they consist of 50% of each enantiomer), but life on Earth primarily (but not exclusively) uses *L*-form amino acids[64,65]. However, over time, non-racemic mixtures can transform into racemic mixtures via a process known as racemization.

Amino acid racemization rates are modelled following Cohen & Chyba (2000)[66], in which the D/L ratio at time *t*+*dt* is given by:

$$\left(\frac{D}{L}\right)_{t+dt} = \frac{e^{2k(t+dt)}R_t - 1}{e^{2k(t+dt)}R_t + 1} \qquad (12),$$

where:

$$R_t = \frac{1 + \left(\frac{D}{L}\right)_t}{1 - \left(\frac{D}{L}\right)_t} \qquad (13),$$

and *k* is the rate constant of the racemization, a function of temperature. Equations ( 12 ) and ( 13 ) allow calculation of how the (D/L) ratio changes with time for different amino acids at different temperatures from any arbitrary initial (D/L) ratio.

Values for *k* were taken from Cohen and Chyba (2000)[66] ($k_{min}$, corresponding to glutamic acid) and Lever et al. (2015)[67] ($k_{max}$, corresponding to a mean proteinogenic amino acid) and in these cases are only available as a function of temperature. Between 0°C and 120°C, $k_{max}$ is always greater than $k_{min}$[13]. To evaluate endmembers of racemization rate, $k_{max}$was used as an upper limit and $k_{min}$ was used as a lower limit.

## Acknowledgements

The authors thank Christopher Glein for helpful and constructive discussions that contributed to model development. This study was financially supported by the Natural Sciences and Engineering Research Council of Canada Discovery and New Frontiers in Research Fund grants to BSL and Natural Sciences and Engineering Research Council of Canada Discovery to OW. The study was also supported by a joint Kavli Foundation and Laukien Science Foundation postdoctoral fellowship to PMH within the Origins of Life Initiative at Harvard University. LMF was supported by NASA FINESST award number 80NSSC21K194. BSL is a Director and OW is a Azrieli Global Scholar of the CIFAR Earth 4D Subsurface Science and Exploration program. The authors thank two anonymous peer reviewers for their constructive comments.

## Code Availability

The code used for the Enceladus abiotic baseline framework and to generate the results presented in this manuscript is available on GitHub: https://github.com/pmhiggins/NutMEG-Implementations. Note for editors and reviewers: please navigate to the IsotopeCycling branch to access the code. It will be merged with the master branch and archived to zenodo after the review process. This custom code uses the NutMEG python package[74] and an existing Enceladus biogeochemical model[3,4].

## Figures and Tables

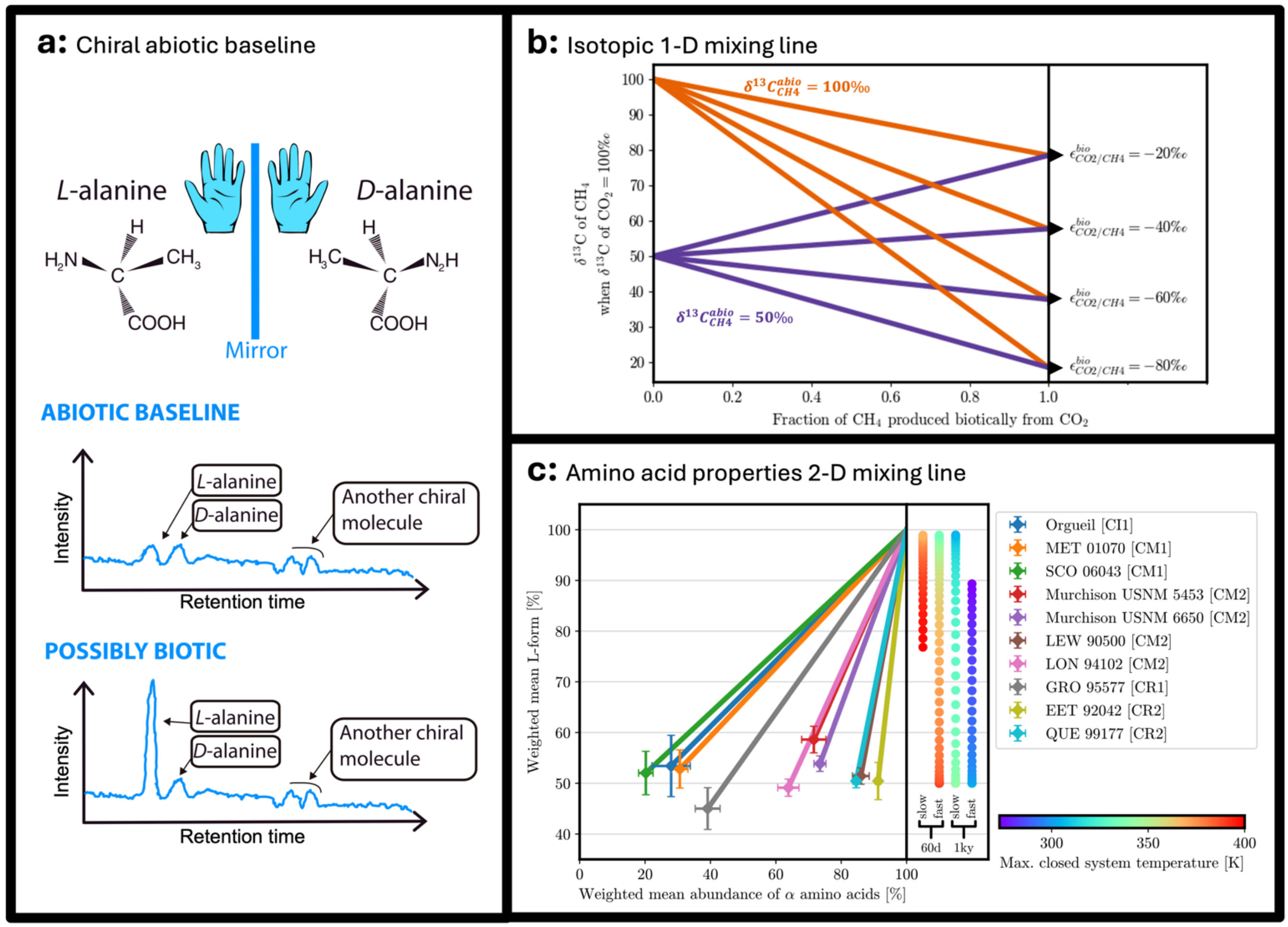

**Fig. 1. Examples of how to interpret biosignatures against an abiotic baseline. a)** An enantiomeric excess (prevalence of one chiral isomer over another, in this case L-alanine over D-alanine). **b)** Illustrative example of a 1D mixing line between abiotic and biotic $CH_4$. Colours show two different abiotic $\delta^{13}C_{CH4}$ contributions while keeping the feedstock $\delta^{13}C_{CO2}$ constant at 100‰. Each mixing line splinters in four directions dependent on microbial isotope enrichment factor $\epsilon^{bio}_{CO2/CH4}$, taken from a characteristic biotic $CH_4$ range. For the $\delta^{13}C_{CH4}$ = 50‰ line it is difficult to disentangle how much $CH_4$ is biotic vs abiotic. **c)** Example of mixing lines that can be drawn between two biosignatures–in this case chiral excess on the vertical axis and abundance of alpha amino acids on the horizontal axis. The biotic endmember is 100% in each, but chirality decays over time and faster at elevated temperature[66]. Dots to the right show the translation in the horizontal axis expected of the biotic endmember over a period of time at different temperatures and racemization rate constants[13]. Data points are for amino acids found in meteorites as possible abiotic endmembers[132].

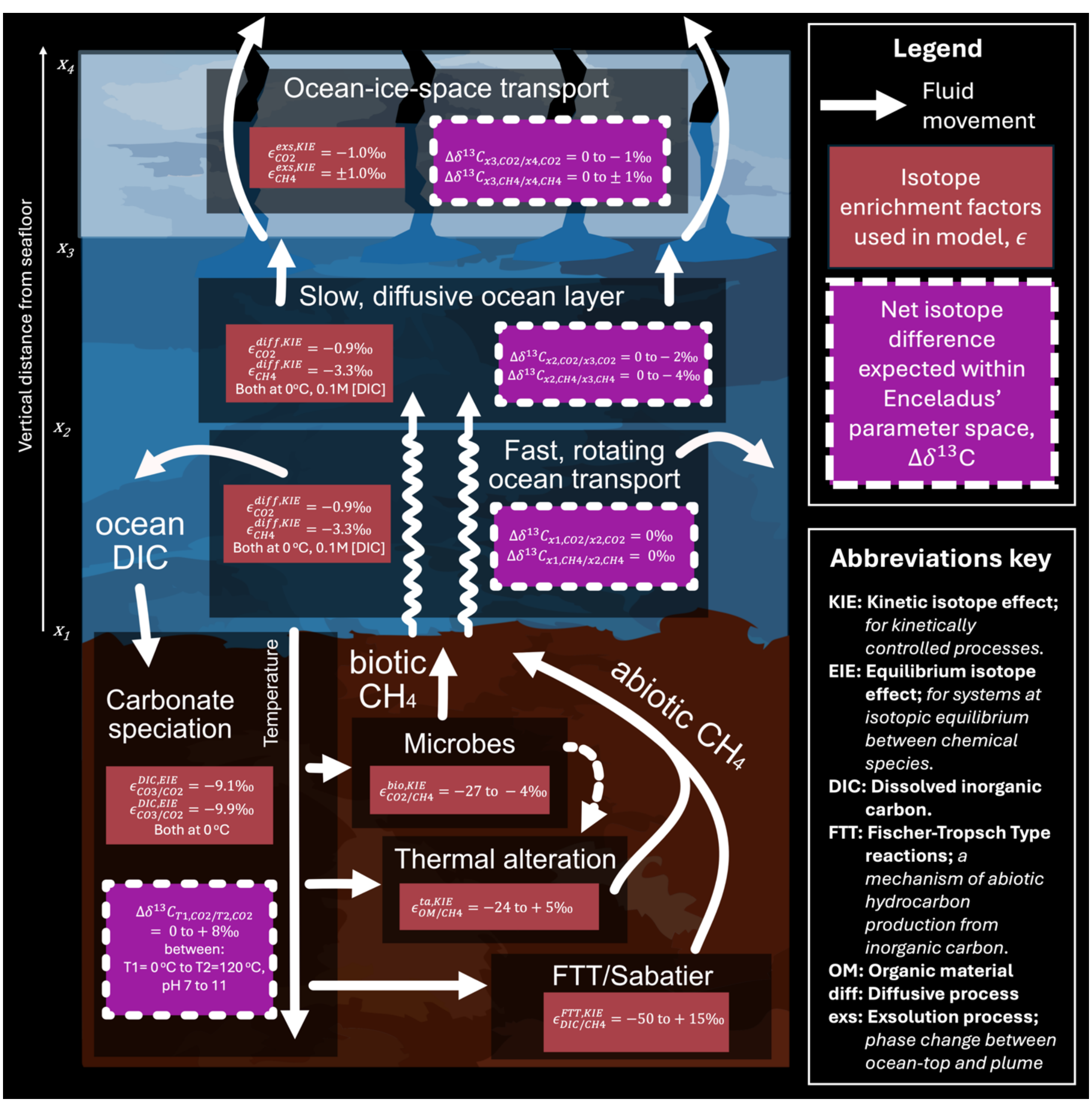

**Fig. 2. Schematic of carbon isotope cycling processes on Enceladus.** Each black box represents a module that monitors key processes affecting influxes and outfluxes of chemical species, and their interactions therein (both bulk chemical and isotopic). For each module, the process-based isotope enrichment factor and/or the model predicted net isotope difference is noted. For isotope definitions refer to **Extended Data Table 1**.

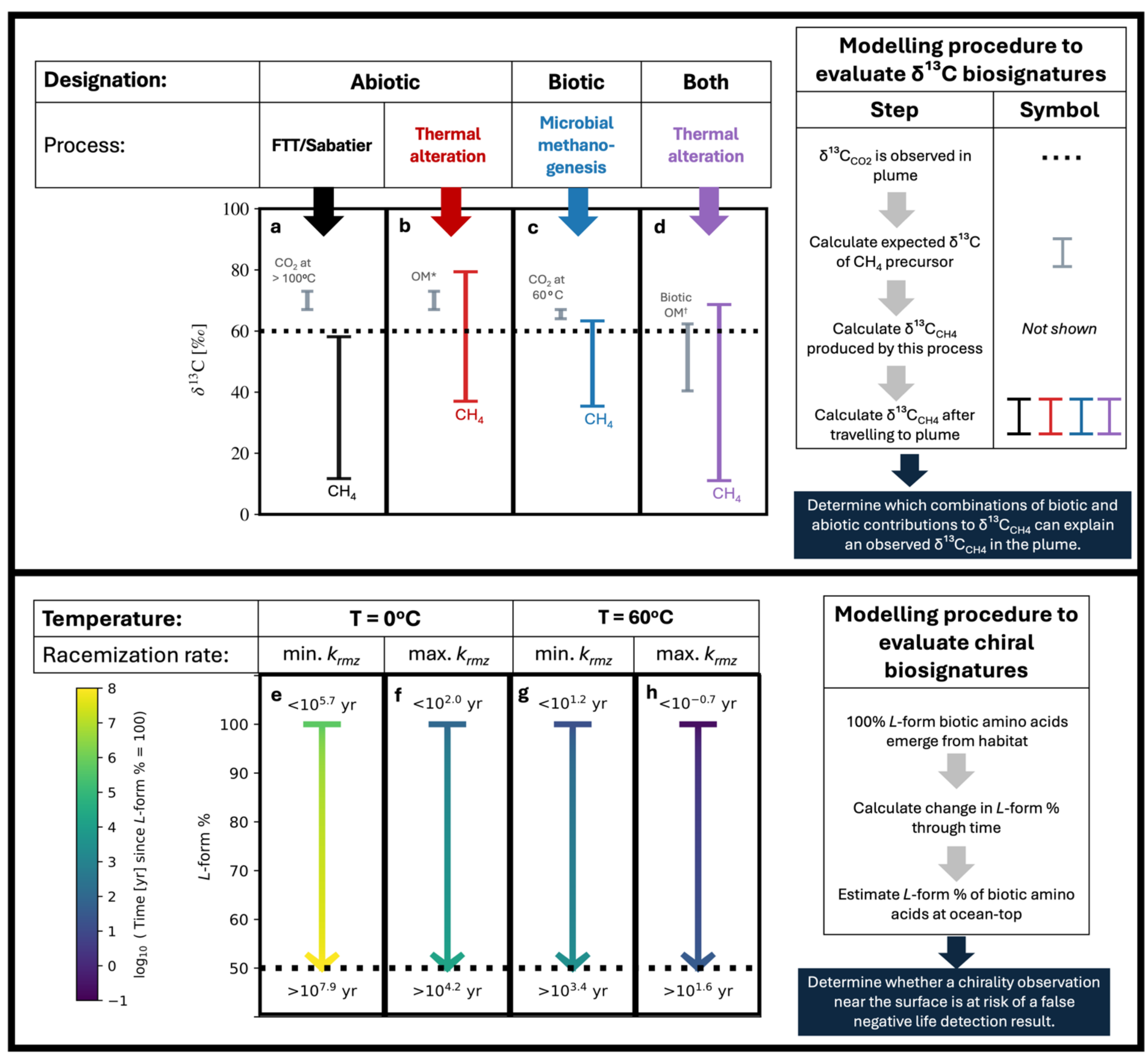

**Fig. 3. Examples of using the framework to evaluate carbon isotope and enantiomeric biosignatures on Enceladus.** *Upper panels***:** Based on a hypothetical observation of $\delta^{13}C_{CO2}$ in Enceladus' space plume of 60 ‰ (dotted line) and the documented range of C-isotope fractionation factors (**Extended Data Fig 1**), these show the estimated end-member $\delta^{13}C_{CH4}$ values in the space plume for $CH_4$ exclusively generated by one of for four possible processes (**a-d**, colored error bars), and the end-member $\delta^{13}C$ values of the $CH_4$ precursor material (gray error bars, with the carbon source

labelled). The four $CH_4$-generating processes are: : **a)** Fischer-Tropsch-Type [FTT] including Sabatier reactions (black), **b)** thermally altered abiotic organic matter (red), **c)** microbial methanogenesis (blue), and **d)** thermally altered biotic organic matter (purple). The hypothetical $\delta^{13}C_{CH4}$ observation could be explained by a linear combination of $\delta^{13}C_{CH4}$ values from each generation process, scaled for their relative $CH_4$ production rates in a simplified case where all four processes are independent of each other (see also the mixing line example in **Fig 1b**). For example, if an observation yielded $\delta^{13}C_{CO2}$=60‰, and $\delta^{13}C_{CH4}$=40‰, any of the four generation processes could have contributed. If $\delta^{13}C_{CO2}$=60‰, and $\delta^{13}C_{CH4}$=20‰, still any of the four generation processes could contribute, but at least one of FTT/Sabatier or thermal alteration of biotic organics must contribute to explain the observation. For the abiotic organic material in panel **b** (*), an equivalent source $\delta^{13}C$ signature to panel **a** is assumed owing to uncertainty in the $\delta^{13}C$ of abiotic organics on Enceladus. For the biotic organic material in panel **d** (†), the source material is biomacromolecules fixed autotrophically from the same carbon source as in panel **c**, using the same isotope enrichment factor as for catabolism. Uncertainties are propagated throughout the ocean (**Fig. 2**) through the analysis described in the main text and Methods. *Lower panels:* For a hypothetical scenario where a 100% biotic mixture of amino acids emerges from Enceladus' seafloor, these show the change in L-form % with time in four different conditions: **e)** slowest racemization rate at 0°C, **f)** fastest racemization rate at 0°C, **g)** slowest racemization rate at 60°C, **h)** fastest racemization rate at 60°C. An L-form % of 50% is consistent with a 100% abiotic mixture. These place bounds on the maximum ocean transport timescale that chiral biosignatures are viable. The maximum timescales decrease with an increasing initial contribution of abiotic amino acids (**Extended Data Fig. 8**). To the right of each row is a flowchart describing how to evaluate the corresponding biosignature evaluation using the framework.

**Table 1. Possible methane ($CH_4$) sources on ocean worlds.**

| Class | Example processes | Carbon source | Notes on Enceladus |
|---|---|---|---|
| Primordial (abiotic) | $CH_4$ trapped in ices of comets, delivered during formation and trapped during differentiation. | n/a (primordial) | Total contribution to present-day composition depends on historic extent of plume activity, Enceladus' building blocks, and volatile loss during differentiation. Some $CH_4$ may be trapped in clathrates[90]. |
| | Thermal alteration* | Refractory organics present in Enceladus' formation material | Requires thermal alteration (see below) to generate light hydrocarbons such as $CH_4$. |
| Ongoing and **does not require** biology (abiotic) | High temperature magmatic $CH_4$ | Carbides and/or $CO/CO_2$ | Unlikely on Enceladus, as requires $T > 500^oC$ and presence of carbides. |
| | Surface-catalyzed C reduction (e.g., Fischer-Tropsch Type or Sabatier reactions) | $CO_2$, CO or HCOOH, w/ metal oxide catalyst | Yields in experiments are usually <<1%, suggesting that long reaction times are required in nature[32]. May occur over geological time scales in fluid inclusions and be liberated in hydrothermal alteration[108]. Higher yields and more rapid when in gas phase and metal-catalyzed. |
| | Uncatalyzed $CO_2$ reduction | $CO_2$ | Requires $T > 150^oC$ to drive reversible reactions between $CO_2$, HCOOH, CO, $CH_3OH$ and ultimately $CH_4$. The $CH_4$ yield is highly context dependent[101,103]. |
| | Postmagmatic high temperature reactions | $CO_2$, metal oxides | Requires $200^oC < T < 500^oC$ and high pressures [refs. [32,84]; references therein]. |
| | High temperature reduction of carbonate minerals | Metal carbonates with graphite/$H_2O/H_2$ | Typically requires $T > 250^oC$ [ref. [84]; references. therein], uncertainties in Enceladus core composition limit applicability at present. |
| | Thermal alteration* | Abiotic organic matter | Hydrothermal pyrolysis of primordial or otherwise abiotically-derived organic material. |
| Ongoing and **requires** biology (biotic) | Thermal alteration | Biotic organic matter | Hydrothermal pyrolysis of biotic organics. $T > \sim 50^oC$-$150^oC$ at high pressure to yield $CH_4$[41]. |
| | Microbial methanogenesis | DIC or DOC | DIC may be delivered to habitat via ocean, water-rock interaction or through syntrophic metabolic pathways (e.g. fermentation). Biology can also act as a sink of $CH_4$ (e.g., methanotrophy). |

**Tab. 1. Footnote:** *$CH_4$ formed by thermal alteration on Earth typically utilizes biotic organic matter as the carbon source. As icy moons appear to have abundant abiotic organic matter, there is possibility for thermogenic production of $CH_4$ from an abiotic source in these settings.

**Extended Data Table 1. Glossary of isotope notation used in this article.**

| Term | Notation | Definition | Additional notes |
|---|---|---|---|
| **Carbon isotope ratio** | $R_X$ | The relative abundance of heavy to light isotopes in chemical species *X*. $R_X = \frac{{C_X}^h}{{C_X}^l}$, where $C_X$ is an abundance measure (*e.g.,* for species containing a single C atom, molality, molarity, or number of moles can be used); *h* and *l* refer to heavy or light isotopologues (for carbon, $^{13}C$ and $^{12}C$ respectively). | |
| **Carbon isotopic composition** | $\delta^{13}C_X$ | Notation for describing carbon isotope composition, expressed as $\delta^{13}C_X(‰) = \frac{R_X - R_{VPDB}}{R_{VPDB}} \times 1000$, where $R_{VPDB}$ is carbon isotope ratio in the Vienna Pee Dee Belemnite standard. | |
| **Isotope-specific rate constant** | $k^{n,j}$ | Rate constant of a process *n* for isotopologue *j*. | |
| **Isotope enrichment factor** | $\varepsilon_{X/Y}^{n,KIE}$ | Isotope enrichment factor for **kinetic-controlled process** n, in which X→Y. Equivalent to $k^{n,h}/k^{n,l} - 1$, where *k* is defined above. | These mechanistic properties are inputs to the modules described in the text and visualised in Fig. 2. They are factors collectively affecting $\delta^{13}C$ due to the concurrence of multiple processes in nature. |
| | $\varepsilon_{X/Y}^{n,EIE}$ | Isotope enrichment factor for process *n* **at equilibrium** $X \rightleftharpoons Y$ . Equivalent to $\frac{R_X}{R_Y} - 1$, where *R* is defined above. | |
| **Net isotope difference** | $\Delta\delta^{13}C_{L,X/Y}$ | The difference in $\delta^{13}C$ between **chemical species** *X* and chemical species *Y* at location *L*, *i.e.*: $\delta^{13}C_{L,X} - \delta^{13}C_{L,Y}$ | Useful for observations or field results, where multiple processes may have contributed to $\Delta\delta^{13}C$. In this work, the net isotope difference is used to show model predictions of how $\delta^{13}C$ can vary depending on location and ongoing processes throughout Enceladus. It is calculated by mechanistic models based on isotope enrichment factors defined above. |
| | $\Delta\delta^{13}C_{X,A/B}$ | The difference in $\delta^{13}C_X$ between **location** A and location B, for chemical species *X* *i.e.:* $\delta^{13}C_{A,X} - \delta^{13}C_{B,X}$ | |

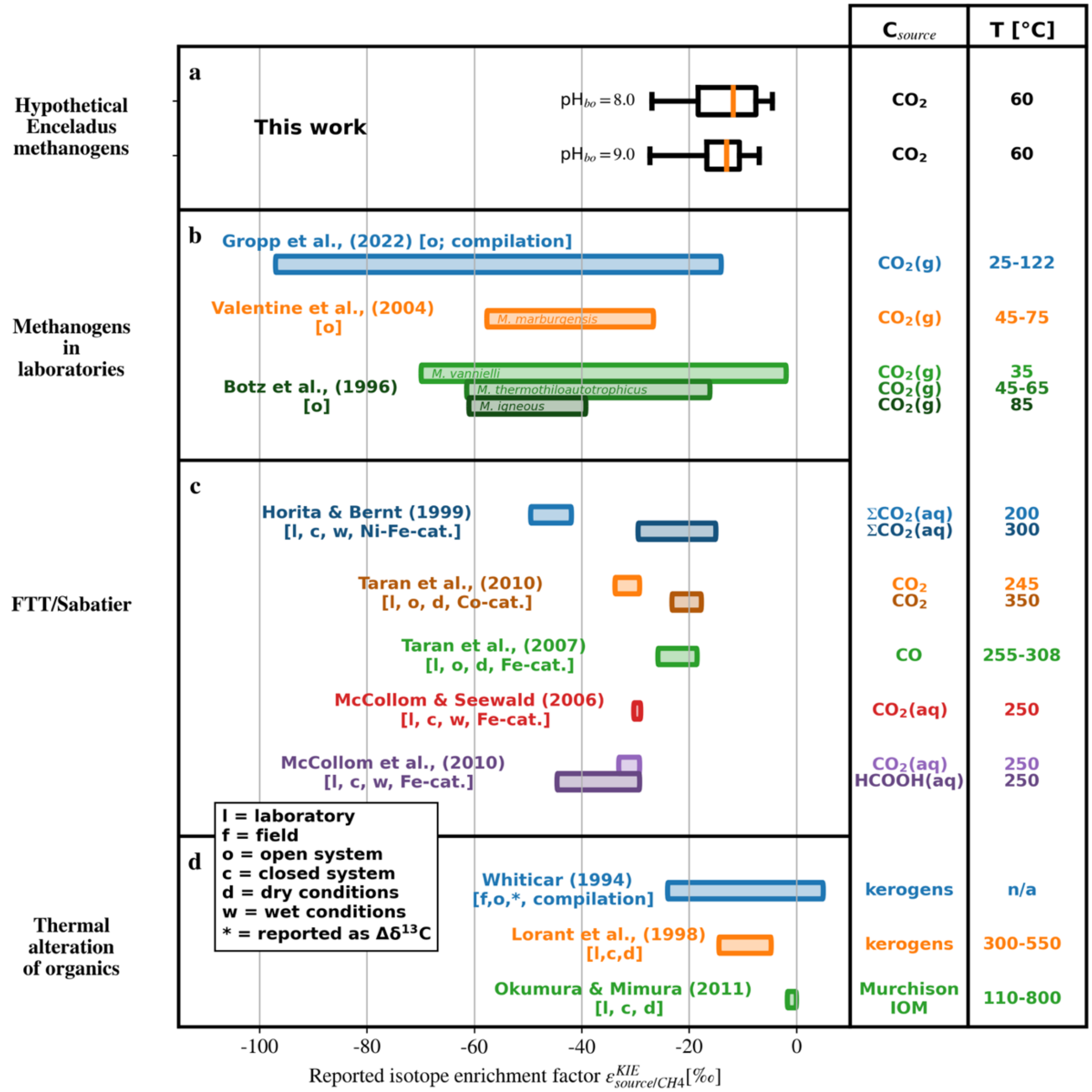

**Extended Data Fig. 1. Compilation of modelled and observed isotope enrichment factors between $CH_4$ and its source carbon**. Panel **a** shows the isotope enrichment factors between $CO_2$ and $CH_4$ for methanogens in an Enceladus seafloor habitat at 60°C. This temperature is used because empirically determined enrichment factors across Enceladus-like $CO_2$ concentrations and methanogenesis affinities are available[56] (see also **Extended Data Fig. 7**). Panel **b** shows species and context-specific isotope enrichment factors of microbial methanogenesis reported from laboratory studies[56,57,59]. Panel **c** shows ranges of isotope enrichment factors associated with $CH_4$

production via Fischer-Tropsch-Type reactions (including Sabatier reactions) in laboratory studies under closed, open, dry, wet, catalysed, and uncatalyzed conditions[36–38,40,85]. Panel **d** shows reported isotope enrichment factors associated with thermal alteration of organic material and the $CH_4$ produced, including biotic and abiotic organic sources[41–43]. Entries denoted with a * were reported as net isotope difference instead of an isotope enrichment factor and hence are not necessarily mechanistic. IOM stands for insoluble organic matter.

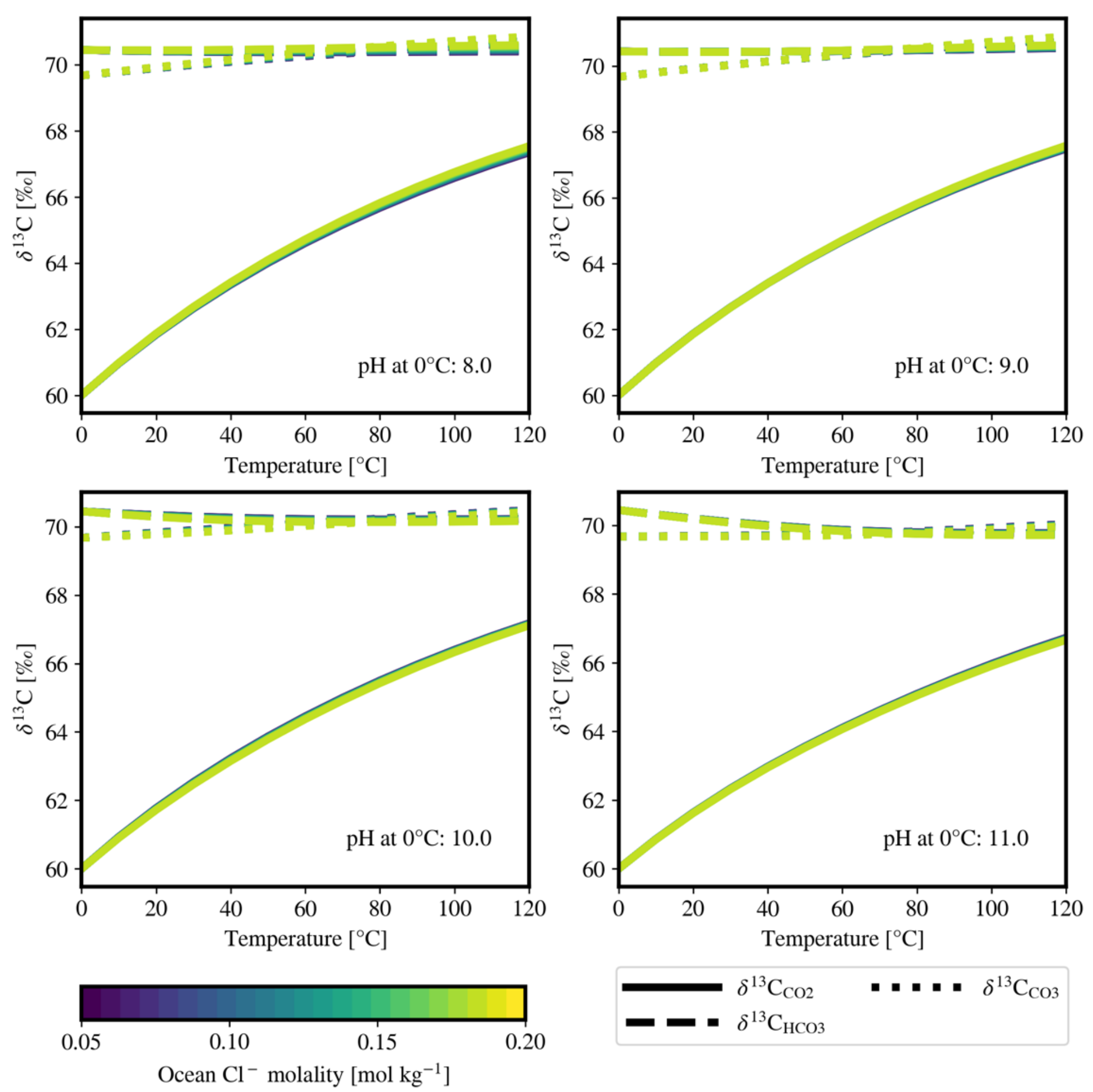

**Extended Data Fig. 2. Example of variation in $\delta^{13}$C for carbonate species upon heating in Enceladus-like conditions.** The nominal $\delta^{13}$C of $CO_2$ inferred by Cassini observations (60‰, notably with uncertainty $\sim \pm$150‰)[25] is chosen for this example. In the Enceladus ocean chemistry model, [$Cl^-$] (indicated by colour) controls the total concentration of dissolved inorganic carbon [DIC]. The carbonate speciation is solved at chemical equilibrium, and a temperature dependent equilibrium isotope effect is applied, using isotope enrichment factors from Deines et al., (1974)[46]. This assumes that all $CO_2$ is dissolved, so without partitioning into $CO_2$(g) or mineral carbonate species, consistent with saturation calculations in Enceladus' ocean[4]. Panels show bulk ocean pH values between 8-11, capturing reasonable uncertainty in Enceladus' pH, show minimal variation in the predicted $\delta^{13}$C for carbonate species. In all four pH values tested, the $\delta^{13}$C of DIC remains constant at approximately 70‰.

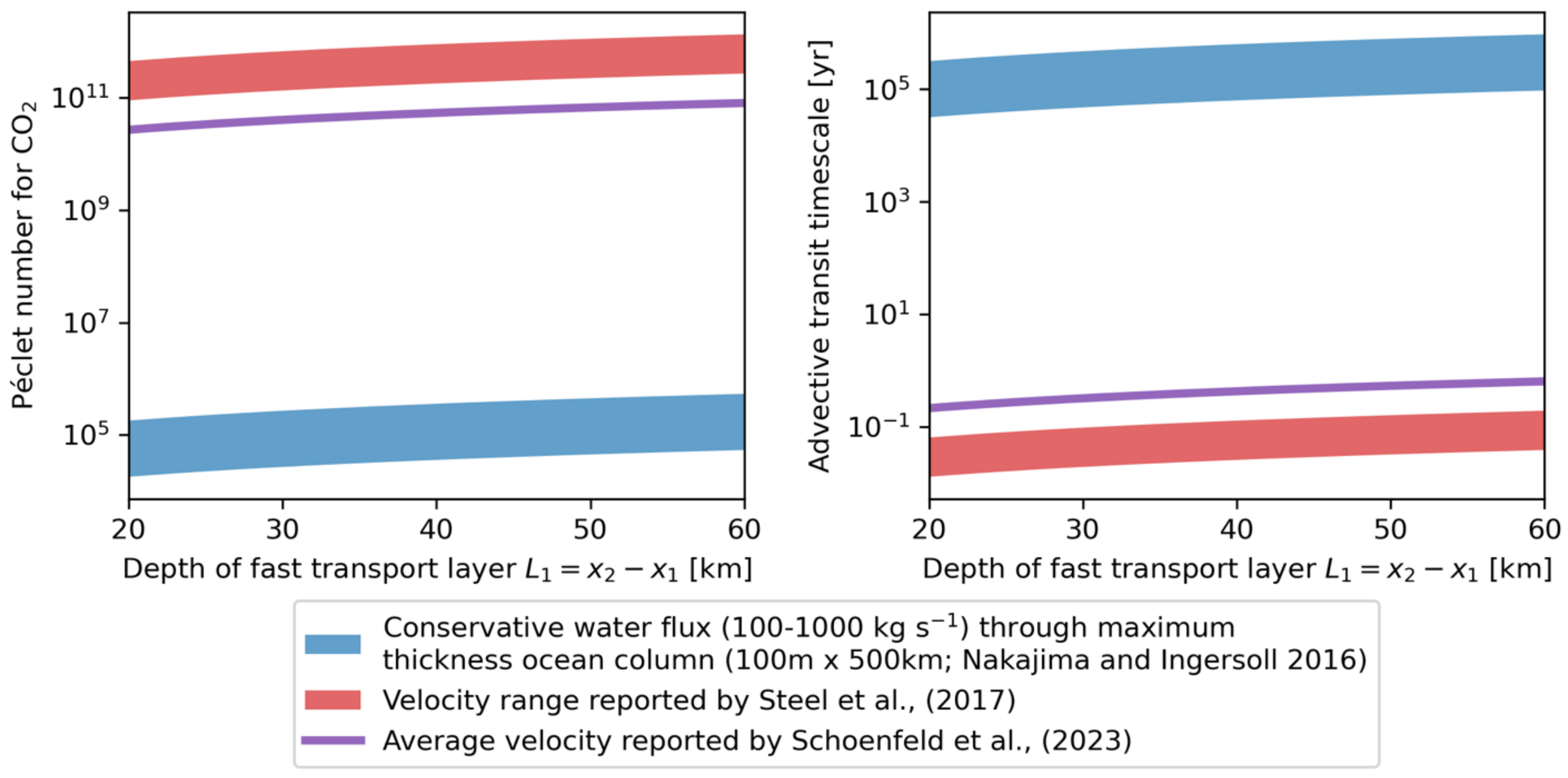

**Extended Data Fig. 3. Estimated Péclet numbers and advective transit timescales in Enceladus' ocean plume**. Calculated using Eq. ( 8 ), and literature velocities from Schoenfeld et al., (2023)[53] (in purple) and Steel et al., (2017)[12] (in red). An estimated lower limit of the Péclet number in Enceladus' ocean plume is shown in blue, using fluid fluxes in the column equivalent to Enceladus' water escape rate (100-1000 kg water per second) coupled with a maximal ascent column width of 100m x 500km (corresponding to the maximum ocean-top area that could feed the space plume from Nakajima and Ingersoll (2016)[119]).

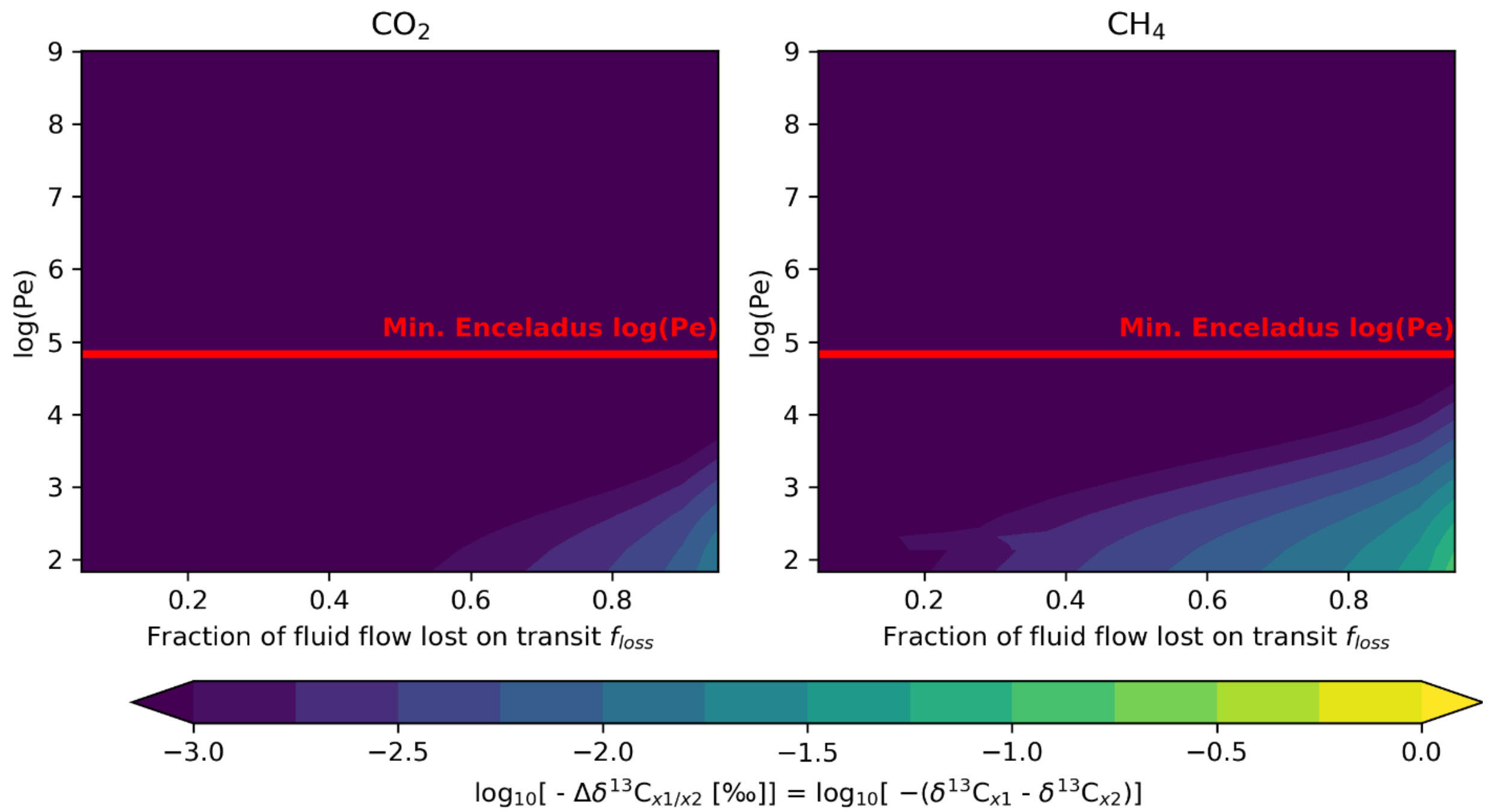

**Extended Data Fig. 4. Maximum change in $\delta^{13}$C of $CH_4$ and $CO_2$ owing to molecular diffusion in Enceladus' fast-moving ocean layer**. Using a suite of input variables in Enceladus-like settings as described in the main text, results were grouped by each simulation's maximum ocean Péclet number ($P_e$) and the fractional loss of flux from the seafloor ($x_1$) to lateral transport ($f_{loss}$). The red horizontal lines show the minimum Péclet number from this work's Enceladus parameter space (**Extended Data Fig. 3**). Areas of the contour below this line are estimated by adjusting the total water mass flow rate to lower values (thus lowering $P_e$), to a smallest value of 0.1 kg s$^{-1}$. A nonzero net isotope difference is only visible–and still less than 1‰–at this velocity. However, if less active worlds have slower water fluxes than Enceladus, diffusion may become important in their isotope cycling even without a stratified, diffusion-dominant layer in their oceans.

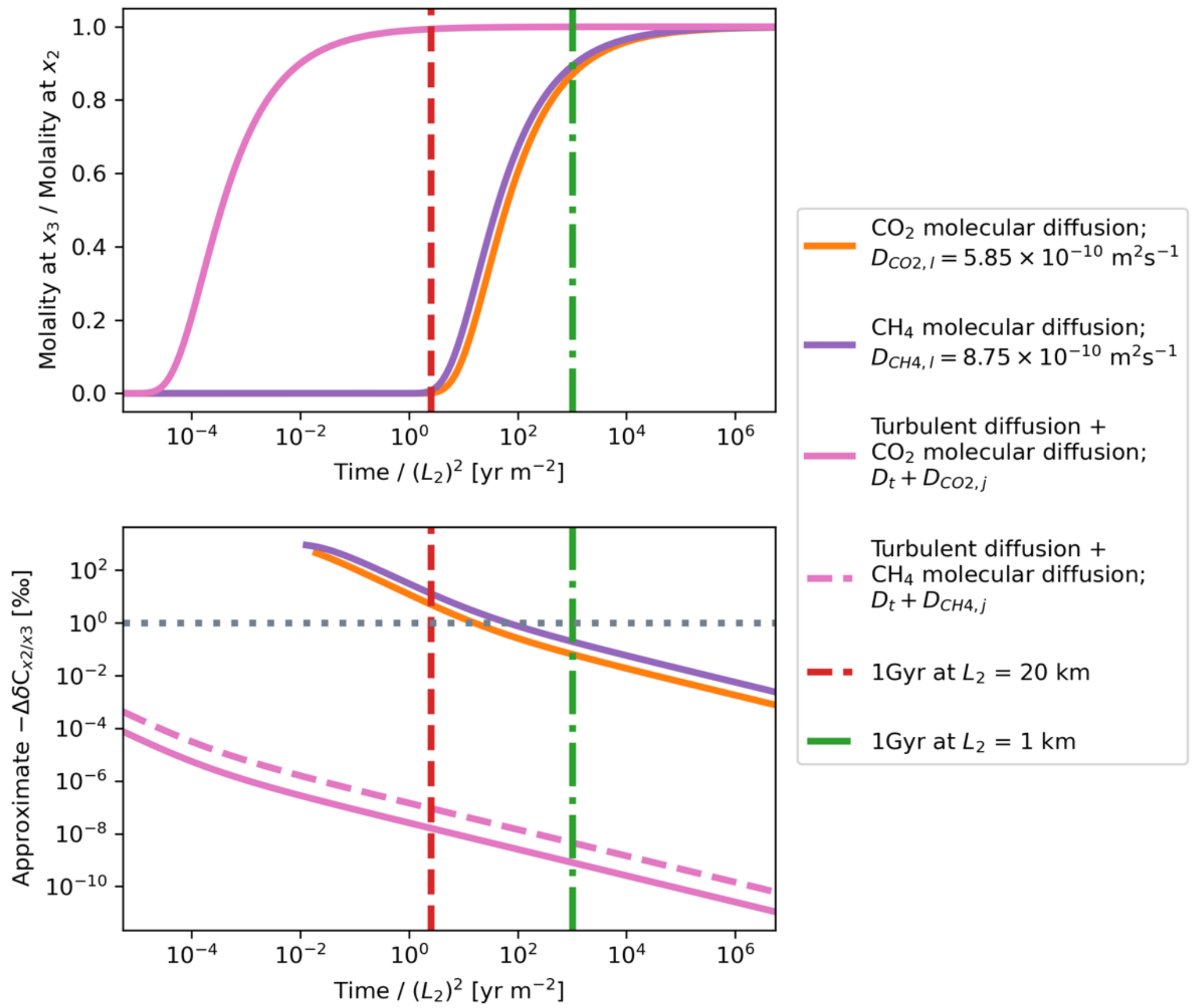

**Extended Data Fig. 5. Bulk concentration profile and net isotopic difference between the bottom ($x_2$) and top ($x_3$) of a stratified layer in Enceladus' ocean where diffusive processes control vertical transport.** Lines correspond to $CO_2$ and $CH_4$ when the only vertical transport mechanism is molecular diffusion (orange and purple respectively) and $CO_2$ and $CH_4$ when turbulent diffusion and molecular diffusion both contribute to vertical transport (pink solid and dashed respectively). The bottom axes show the time since the simulation began divided by layer depth squared: $(L_2)^2 = (x_3 - x_2)^2$, allowing a comparison of different simulation times and hypothetical layer depths. To contextualise the horizontal axes, two examples are shown with vertical lines: after 1Gyr with a 20km thick layer in red, and after 1 Gyr with a 1km thick layer in green. *Top panel*: Ratio between bulk concentration at top of layer $x_3$ and bottom of layer $x_2$ to show the timescale associated with the $x_2$ composition reaching $x_3$. *Bottom panel:* Predicted net isotope difference between $x_2$ and $x_3$, approximated as $[R_{x2}/R_{x3} - 1]$ as described in the Methods. The horizontal dotted line shows where a net isotope difference of 1‰ would be.

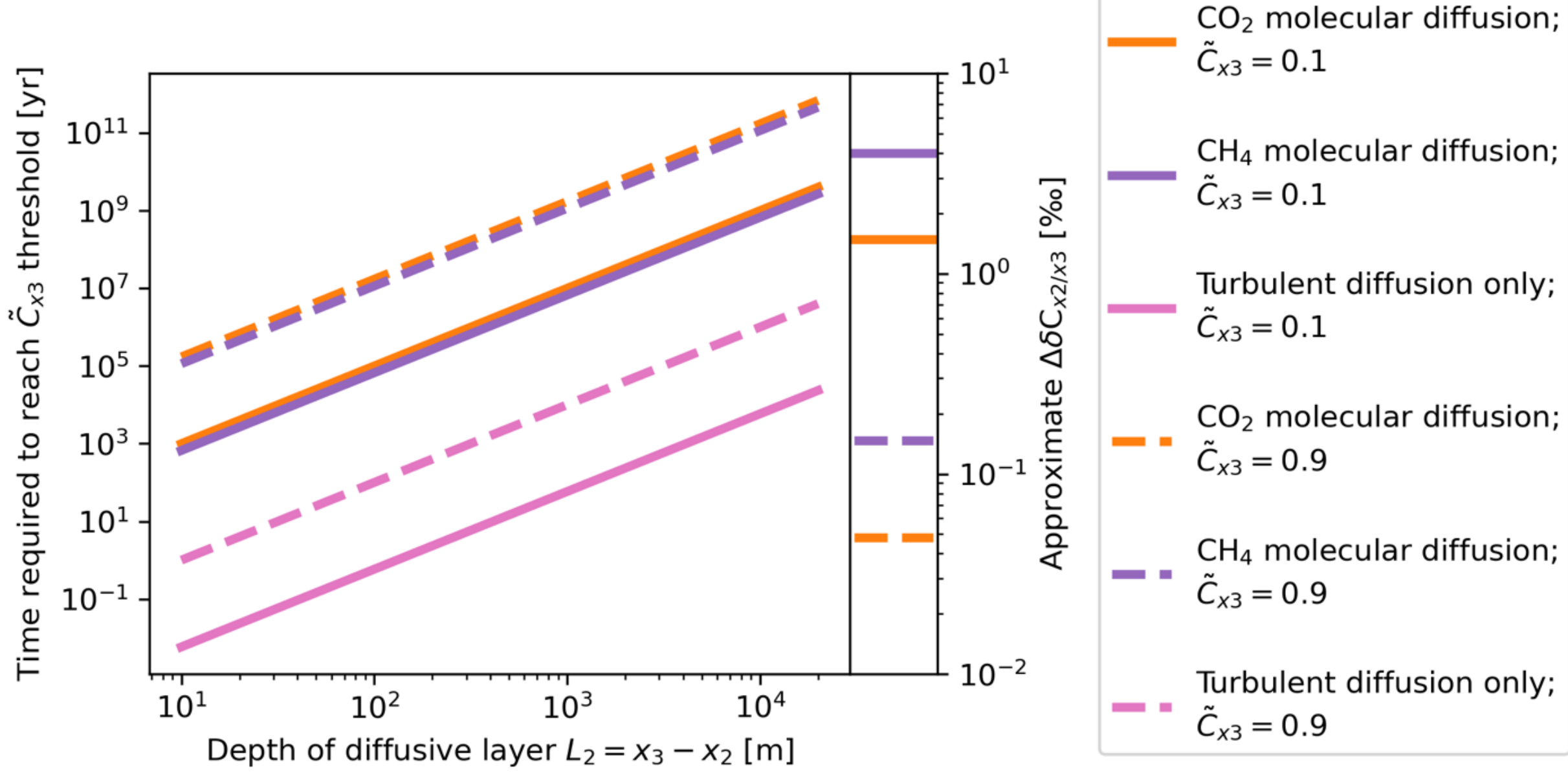

**Extended Data Fig. 6. Predicted timescales of transport and net isotope difference across a stratified layer in Enceladus' ocean where diffusive processes control vertical transport.** Lines correspond to $CO_2$ and $CH_4$ when the only vertical transport mechanism is molecular diffusion (orange and purple respectively) and when turbulent diffusion is the primary contributor to vertical transport (pink). Solid lines are the timescale and net isotope difference at the point when the concentration at the top of the layer $\widetilde{C_{x3}}$ is 10% that at the bottom of the layer (location $x_2$). Dashed lines are the timescale and net isotope difference when the concentration at $x_3$ is 90% of its value at $x_2$.

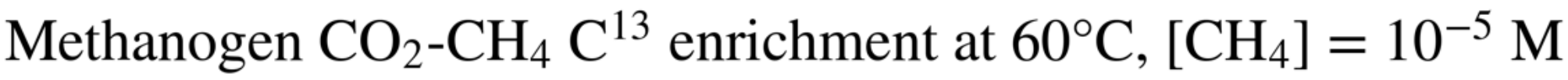

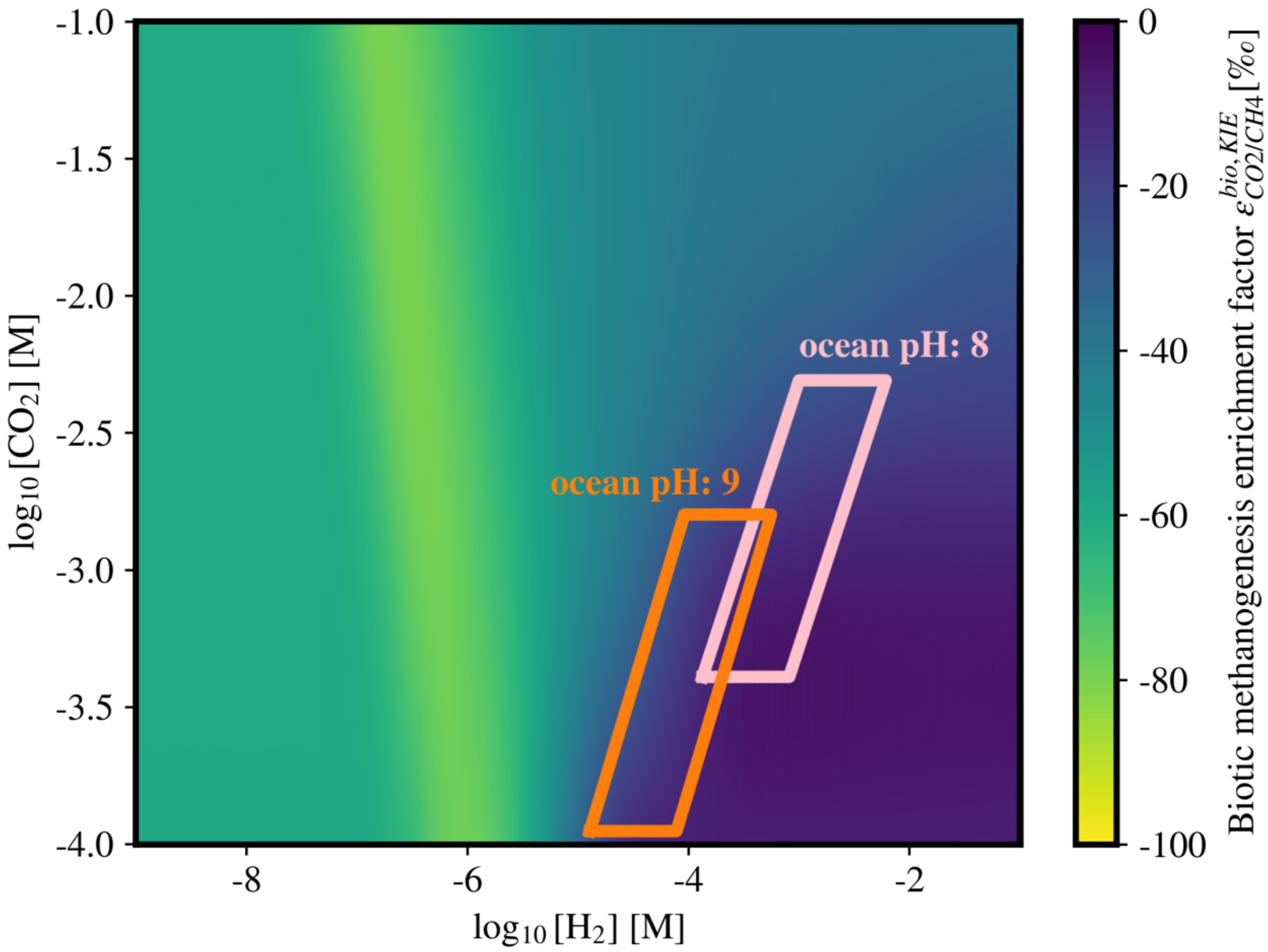

**Extended Data Fig. 7. Biotic methanogenesis enrichment factor between $CO_2$ and $CH_4$ on Enceladus.** The enrichment factor is shown as a function of habitat [$CO_2$] and [$H_2$] based on the model by Gropp et al., (2022)[56]. Boxes capture the range of [$H_2$] and [$CO_2$] expected in an Enceladus seafloor habitat at 60°C with a 0°C ocean pH of 8 and 9[4].

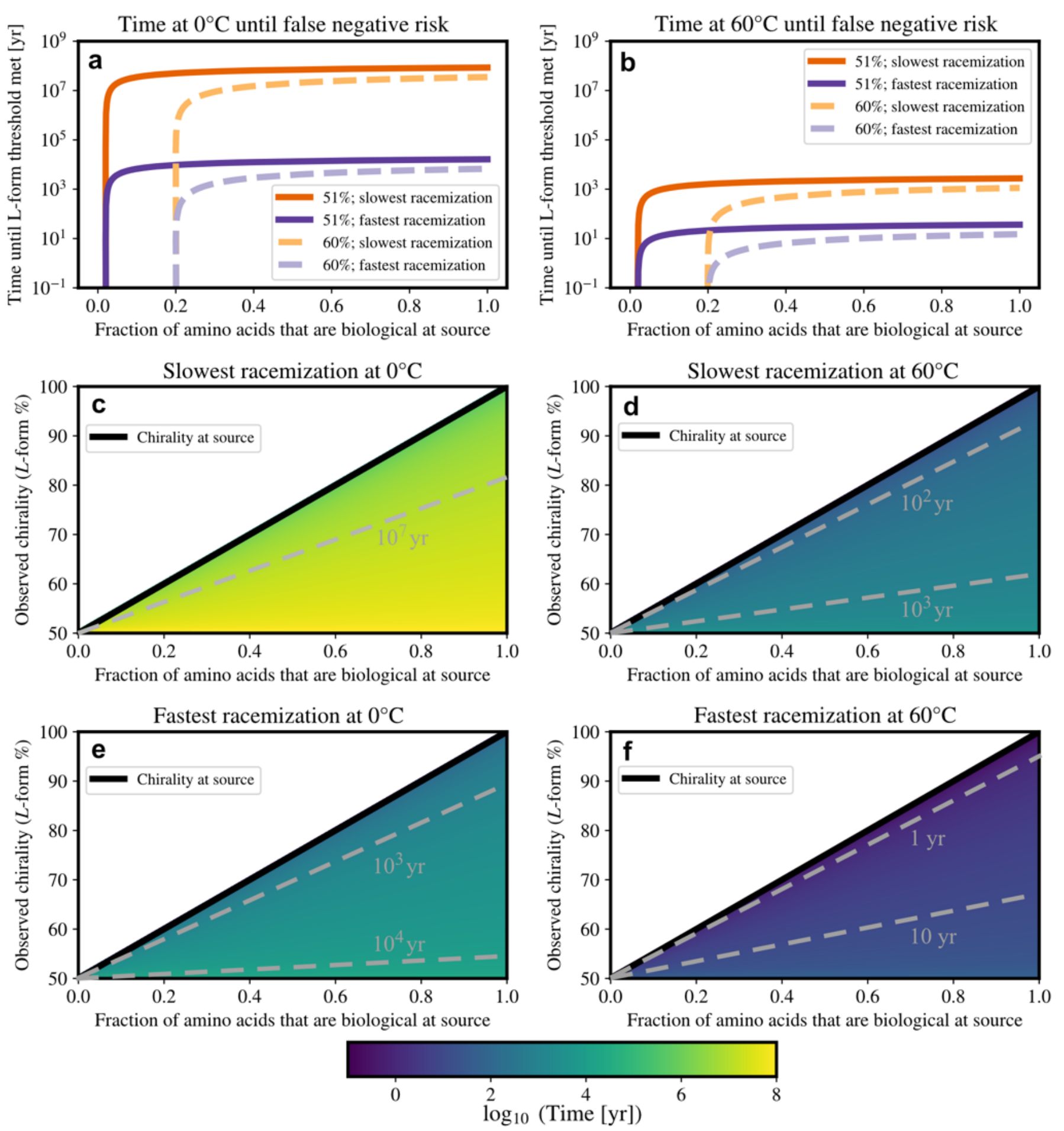

**Extended Data Fig. 8. Racemization timescales of mixed solutions of biotic and abiotic amino acids.** Panels **a** and **b** show the time until a threshold observed L-form % is reached at 0°C and 60°C respectively. In both panels, dashed lines correspond to reaching L-form % = 60, and solid lines correspond to reaching L-form % = 51. Orange lines are for the slowest anticipated racemization rate constant, and purple lines are for the fastest anticipated racemization rate constant. Panels **c-f** show the modelled observed L-form % (vertical axes) as a function of the fraction of amino acids that are biological at source (horizontal axes) and the time since they left the source (indicated by colour). Dashed grey lines highlight the specific relationship at the labelled times. The four panels show the four racemization case studies: **c)** slowest racemization rate at 0°C, **d)** slowest racemization rate at 60°C, **e)** fastest racemization rate at 0°C, **f)** fastest racemization rate at 60°C.